\documentclass{aa}  

\usepackage{graphicx}
\usepackage{txfonts}
\usepackage{xcolor}
\usepackage{soul}
\usepackage{txfonts}
\usepackage[colorlinks=true,citecolor=blue,urlcolor=blue]{hyperref}
\usepackage{breakurl}
\begin{document}

   \title{The role of radial migration in open cluster and field star populations with {\em Gaia} {\sc dr3}}


 \titlerunning{Radial migration in stellar populations} 
\authorrunning{Viscasillas V{\'a}zquez et al.}
   
 \author{
C. Viscasillas V{\'a}zquez\inst{\ref{vilnius}}\and 
L. Magrini\inst{\ref{oaa}} \and 
L. Spina\inst{\ref{oapd}} \and 
G. Tautvai{\v s}ien{\. e}\inst{\ref{vilnius}} \and 
M. Van der Swaelmen\inst{\ref{oaa}} \and 
S. Randich\inst{\ref{oaa}} \and
G.~G. Sacco\inst{\ref{oaa}} }

\institute{
Institute of Theoretical Physics and Astronomy, Vilnius University, Sauletekio av. 3, 10257 Vilnius, Lithuania. \label{vilnius}
\and 
INAF - Osservatorio Astrofisico di Arcetri, Largo E. Fermi 5, 50125, Firenze, Italy.  \label{oaa} 
\and
INAF - Padova Observatory, Vicolo dell'Osservatorio 5, 35122 Padova, Italy\label{oapd} 
}

   \date{Received 22 May 2022/ Accepted 28 September 2023}

 
  \abstract
    {The survival time of a star cluster depends on its total mass, density, and thus  size, as well as on the environment in which it was born and in which lies. Its dynamical evolution is influenced by various factors such as gravitational effects of the Galactic bar, spiral structures, and molecular clouds. Overall, the factors that determine the longevity of a cluster are complex and not fully understood.}
   {This study aims to investigate if open clusters and field stars respond differently to the perturbations that cause radial migration. In particular, we aim at understanding the nature of the oldest surviving  clusters. }
   {We compared the time evolution of the kinematic properties of two {\em Gaia} {\sc DR3} samples: the first sample is composed of $\sim$40 open clusters and the second one of $\sim$66,000 MSTO field stars. Both selected samples are composed of stars selected with the same quality criterion, belonging  to the thin disc, in a similar metallicity range, located in the same Galactocentric region [7.5-9~kpc] and  with ages >1 Gyr. We performed a statistical analysis comparing the properties of the samples of field stars and of open clusters.}
   {A qualitative comparison of kinematic and orbital properties reveals that clusters younger than 2-3 Gyr are more resistant to perturbations than field stars and they move along quasi-circular orbits. Conversely, clusters older than approximately 3 Gyr have more eccentric and inclined orbits than isolated stars in the same age range. Such orbits lead them to reach higher elevations on the Galactic plane, maximising their probability to survive several Gyr longer. A formal statistical analysis reveals that there are differences among the time evolution of most of the kinematic and orbital properties of field stars and open clusters. However, the comparison between some  properties (e.g. V$_{\phi}$ and $L_{Z}$) do not reach a sufficient statistical significance. }
   {Our results suggest that oldest survived clusters are usually more massive and move on orbits with higher eccentricity.  Although they are still reliable tracers of the Galaxy's past composition, they do not reflect the composition of the place where they are currently found. Therefore, we cannot avoid considering kinematic properties when comparing data and models of chemical evolution, taking also into account the intrinsic differences between clusters and isolated stars. To validate the results, new studies that increase the sample of open clusters, especially at older ages, are needed.}


   \maketitle
%

\section{Introduction}
\label{section:introduction}

Radial migration is due to interactions of stars with spiral arms or other non-axisymmetic structures in the Galactic potential \citep{Sellwood_2002}. It produces changes in  stellar orbits, which are initially circular. This results in a radial displacement of the stars with respect to the Galactocentric radius (R$_{\rm GC}$) at which they formed. 

Since migration redistributes  stellar populations to different parts of the Galactic disc, the abundances measured now in stars of different ages in a given Galactic location cannot be considered as completely representative of the past interstellar medium composition in that place. 
It is indeed necessary  considering the effect of radial migration for a comprehensive understanding  of Galactic chemical evolution \citep[e.g.][]{Kubryk_2013}. For instance, \citet{Loebman_2016} found that radial migration has a significant impact on the shape and width of the metallicity distribution functions (MDFs) at different Galactocentric distances. 

The study of the chemo-dynamical properties of open clusters and field stars can provide valuable information about the impact of radial migration in the Milky Way disc. Open clusters are groups of coeval stars that formed together from the same molecular cloud, sharing the same chemical composition \citep[see][for a review about star cluster formation and evolution in galactic and cosmological contexts]{Renaud18}. Since stars in clusters are gravitationally bound, they move together in the Galactic potential field and are subject to the same perturbations. Therefore, open clusters are expected to migrate as a coherent group. However, some perturbations might also cause individual stars to escape from the cluster \citep[see][for numerical simulations capturing  the importance of the
small-scale, rapidly varying tidal component in altering the mass-loss of clusters]{li17} and become part of the field population \citep[e.g.][]{Moyano_Loyola_2013} and $vice$ $versa$ \citep[e.g.][]{Mieske_2007}.  \citet{Fukushige_2000} showed the escape time of a star from its parent cluster is related also to orbital parameters. Gravitational perturbations can also lead to cluster-cluster interactions \citep[e.g.][]{Khoperskov_2018,delaFuenteMarcos_2014}, which may be important for the formation of the bar in disc galaxies \citep{Yoon_2019}.

Due to their different initial conditions, stars in open clusters and isolated stars are expected to be  differently affected by migration. 
Both populations are impacted by gravitational interactions with the spiral arms and the Galactic bar, with other clusters and with molecular clouds. However, member stars of open clusters are also strongly influenced by the cluster's internal dynamics and we can consider each cluster, taken as a whole, as a more massive particle than a single star, and therefore we might hypothesise that its kinematics is impacted differently by gravitational interactions with respect to  single stars. N-body simulations  of gravitational interaction with particle of different masses would be needed to assess the amount of these differences. Seminal works, as that of  \citet{Terlevich_1987}, investigated the effect of tidal heading and molecular cloud encounters in shaping the halo of clusters and determining their lifetime. More recent N-body simulation analysed the interactions of  star clusters  with spiral arms \citep{fujii12}.

From an observational point of view, \citet{Spina_2021} using data from the Galactic Archaeology with HERMES \citep[GALAH;][]{DeSilva_2015,Buder_2021} and Apache Point Observatory Galactic Evolution Experiment \citep[APOGEE-1 and APOGEE-2;][]{Ahn_2014,Jonsson_2020} surveys found that the open cluster population traces the distribution of chemical elements differently than field stars. The authors suggested that such a difference is a consequence of selection effects shaping the demography of the two populations in different ways. In fact, while field stars undergoing frequent interactions with the Galactic potential would simply migrate on different orbits, open clusters would also dissipate until they face their complete disruption. The effect of radial migration has been studied on some specific open clusters. A well-known example is the open cluster NGC\,6791, one of the oldest open  and  most metal rich clusters. For this cluster, \citet{Jílkova_2012} proposed a model suggesting its migration from the inner disc to its current location due to a strong influence of the bar and spiral-arm perturbations on its orbit. 
A systematic study of migration in a significant sample of open clusters was carried out by \citet{chen20}. They used a sample of 146 open clusters to investigate the kinematics and metallicity distribution of open clusters in the Galactic disc, and found evidence for significant radial migration.  \citet{Zhang_2021} analysed the metallicity gradient of 225 open clusters identifying three sequences of clusters that represent outward migrators, {\em in situ} clusters, and inward migrators. Their study suggests that radial migration is an important process in the evolution of the Galactic disc and has a complex effect on the metallicity gradient.

Overall, the survival of star clusters is a complex process that depends on a variety of factors. While some clusters may be more susceptible to disruption than others and  the exact conditions that determine their longevity are still not fully understood.
Their structural parameters, such as mass, density, and size, are expected to play a crucial role in determining their survival time \citep{deGrijs_2007}. More massive clusters are generally more tightly bound  \citep{Kruijssen_2012}, and therefore less subject to disruption, while less massive clusters are likely more easily disrupted. In addition, irrespective of their mass, more compact and denser clusters have a higher probability of surviving, as they have a greater gravitational binding energy and are less likely to be disrupted by external forces \citep[see][for a discussion of the cluster compactness as a function of Galactocentric distance]{angelos23}. The environment in which a cluster is located can also affect its survival \citep[e.g.][]{Grebel_2000, Lamers_2005}. Thus, the dissolution mechanisms of the clusters (initial gas loss, stellar evolution, relaxation and external tidal perturbations) change over time and also depend on the position of the cluster in its parent galaxy \citep{Baumgardt_2009}. Clusters that are located in denser regions of the Galaxy, such as the disc towards the Galactic Center or spiral arms, are more likely to be subject to disruptive tidal forces caused by the Bar, the spiral arms  or molecular clouds \citep[e.g.][]{Portegies_2002,Baumgardt_2003,Gieles_2006, Gieles_2007}. On the other hand, clusters that are located in less dense regions, such as the Galactic halo, are usually more isolated and therefore less susceptible to disruption \citep{Meng_2022}. Finally, the dynamical evolution of the cluster can also play a role in its survival. Over time, the cluster will undergo a process of mass segregation, where more massive stars sink to the center of the cluster and interact more strongly  each with the others \citep[e.g.][]{Allison_2010}. In the short term this can lead to the ejection of stars of lower masses from the cluster, and can ultimately cause the cluster to dissolve. However, if the cluster is able to maintain a balance between the processes of mass segregation and two-body relaxation, it may be able to survive for a longer period of time.

By comparing the properties of clusters and field stars, the present paper aims to shed light on the role of radial migration in shaping the distribution in the Galactic disc of its stellar populations and of their metallicity. Furthermore, we aim at clarifying  the nature of  the surviving old clusters, a small number out of the total percentage of known clusters, and whether it is possible to use them as tracers of past  composition of the Galactic disc.
The paper is structured as follows: Section~\ref{section:samples} provides a description of the {\em Gaia} {\sc DR3} open clusters and field stars samples, how the sub-samples were selected, as well as how the ages of the latter were determined. Section~\ref{section:properties} compares the kinematic properties of clusters and field stars, specifically their space velocities, orbits, and actions over time. Finally, Section~\ref{section:discussion} discusses the results and Section~\ref{sec:conclusions} draws our conclusions.





\section{The samples}
\label{section:samples}

\subsection{The sample of open clusters in {\em Gaia} dr3}

\begin{figure}
  \resizebox{\hsize}{!}{\includegraphics{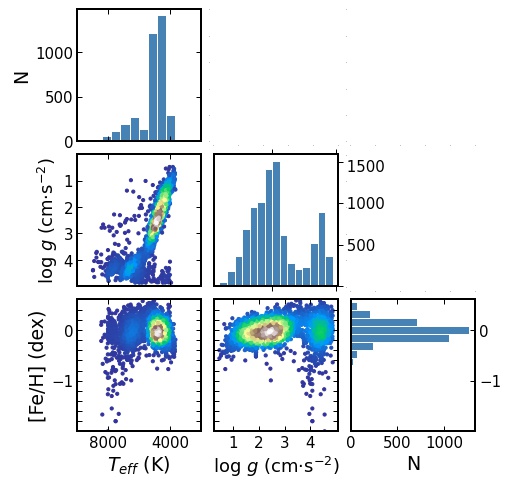}}
  \caption{Relationships of the calibrated GSPspec parameters for the $\sim$4,000 member stars of open clusters with HQ=1 and/or MQ=1 quality flags.}
  \label{fig:3718_pairwisse}
\end{figure}

\begin{figure}
  \resizebox{\hsize}{!}{\includegraphics{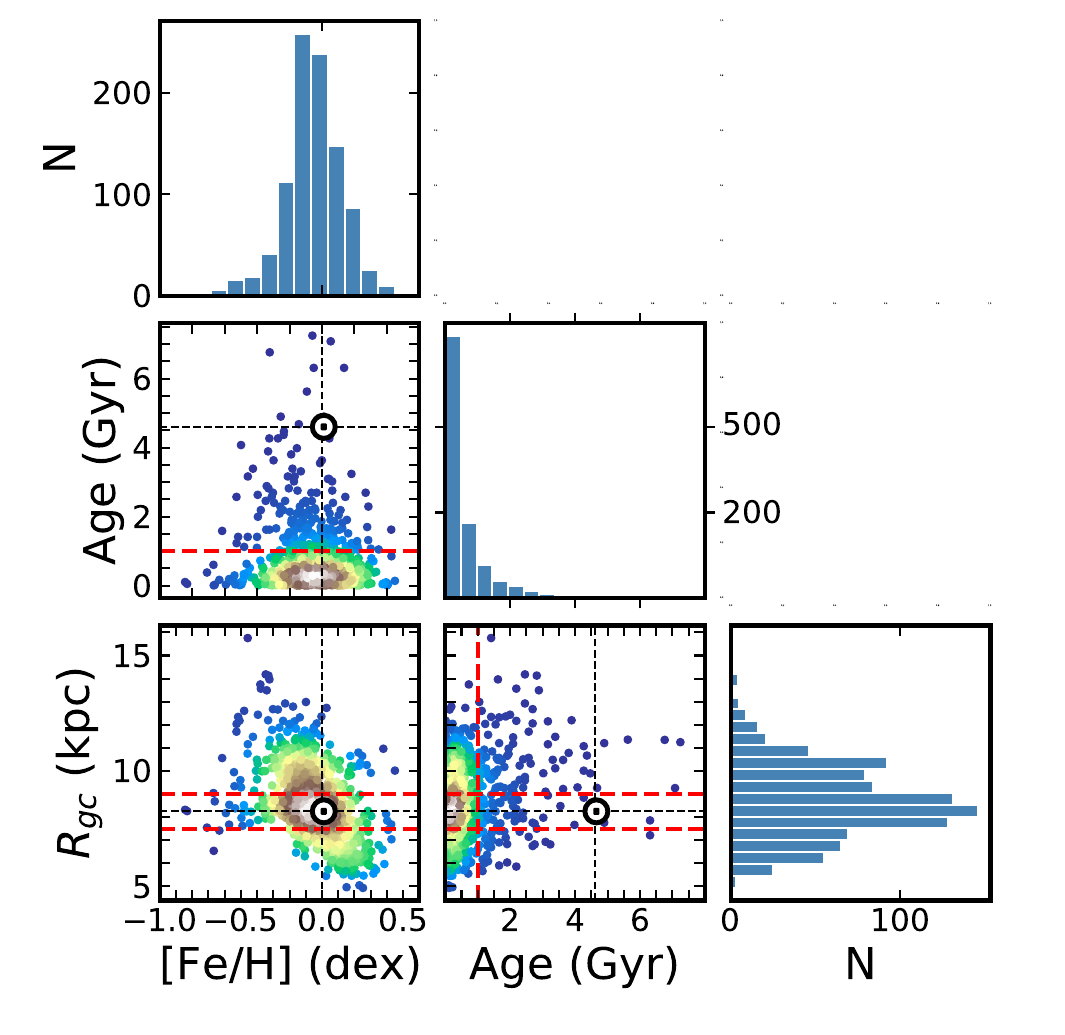}}
  \caption{Relationships of ages and galactocentric distances for the $\sim$1,000 open clusters. The dashed red lines indicate the cutoff in age and R$_{\rm GC}$}.
  \label{fig:998_pairwisse}
\end{figure}

We consider a sample of $\sim$300,000 member stars of $\sim$2700 open clusters with data from {\em Gaia} {\sc DR3} \citep{gaiadr3}, from which we select $\sim$8,000 member stars with available Gaia spectroscopic atmospheric parameters and abundances from the Gaia General Stellar Parametrizer from spectroscopy (GSPspec) \citep{Recio-Blanco_2022}. 
To select a sample of high quality data, we used HQ (High Quality) and MQ (Medium Quality) indicators derived from a combination of {\em Gaia} GSPspec flags and defined in \citet[][see their Appendix B for a complete defiition of the ranges of the used GSPspec flags to produce the HQ and MQ samples]{Gaia23_chemical_cartography}. 
The MQ sample defined in \citet{{Gaia23_chemical_cartography}} contains about $\sim$4,100,0000 stars with median uncertainty in [M/H] of about 0.06~dex and  median uncertainty in [$\alpha$/Fe] of about  0.04 dex, while the HQ sample stars ($\sim$2,200,000) have with very low parameter uncertainties, in particular a median uncertainty in [M/H] $\sim$0.03 dex and $\sim$0.015 in [$\alpha$/Fe]. Through this selection we obtain a sample of about $\sim$4,000 members of open clusters. The relationships between the calibrated GSPspec parameters ([Fe/H], $T_{\mathrm{eff}}$, log$g$) for our sample of open cluster member star, belonging to about $\sim$1,000 OCs, are shown in Fig.~\ref{fig:3718_pairwisse}. 
Since the log$g$ determination is slightly biased in Gaia GSP-Spec, we use the calibrated values {\em logg\_gspspec\_calibrated}, \textit{mh\_gspspec\_calibrated} and \textit{alphafe\_gspspec\_calibrated} as presented in \citet[][Sect 9.1.1 and 9.1.2, Eqs. 1, 2 and 5]{Recio-Blanco_2022}. These corrections basically use fitted coefficients from literature trends to adjust log$g$, and a similar correction is suggested for metallicity and [$\alpha$/Fe] based on a fourth-degree polynomial fit of residuals against uncalibrated log$g$.
The membership of stars in clusters, as well as general cluster parameters and their ages  are taken from \citet{CantatGaudin20}. Relationships between the average metallicity, age and Galactocentric distance for our sample of open cluster are shown in Fig.~\ref{fig:998_pairwisse}. 


To analyse the effect of migration, we consider only cluster with an age $\ge$1~Gyr. Indeed, the orbits of the youngest open clusters are not expected to be strongly affected by migration given the limited number of encounters or disturbances they may have had in their short lives. This reduces the sample to 201 clusters (20\% of the total), from which we extract those with high probability (P>0.9) to belong to the thin disc which is estimated as indicated below.  

\begin{figure}
  \resizebox{\hsize}{!}{\includegraphics{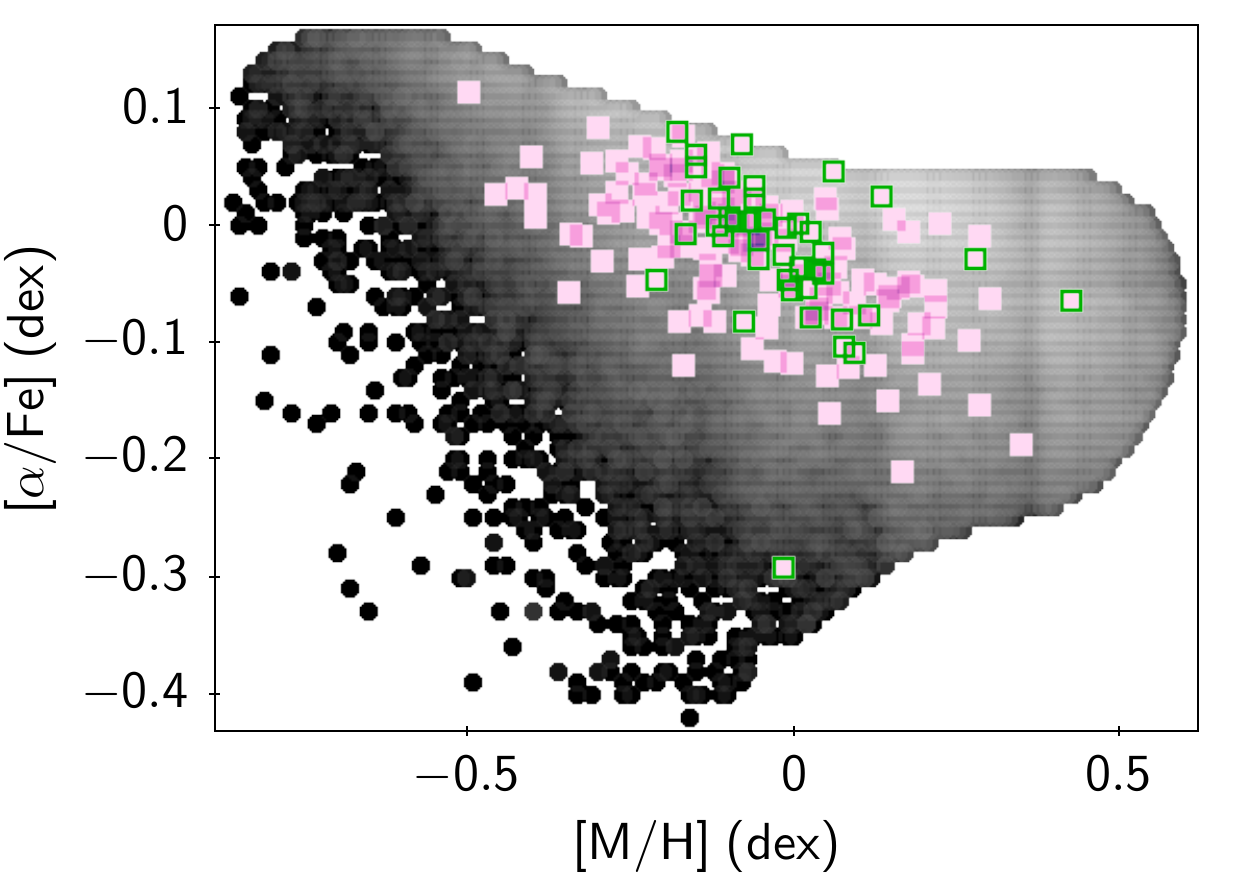}}
  \caption{TWD for the $\sim$200,000 MSTO field stars (grey symbols) and $\sim$170 open clusters (pink symbols), both potentially from thin disc and with similar characteristics. Clusters with a green edge represent those that are in the solar region ($\sim$40 open clusters).}
  \label{fig:alphafe_ocs_field_thin}
\end{figure}

At this scope, we compute a thin-thick disc components separation and membership probability using the Support Vector Machines (SVMs) analysis \citep{Boser92}, already adopted in \citet{Viscasillas_2022}. We defined a training set based on the sample of \citet{Costa_Silva20}, which have similar characteristics as our sample, with [$\alpha$/Fe]-[Fe/H] derived by \citet{DelgadoMena17}. We included the thin and thick disc populations, as well a high-$\alpha$ metal-rich population (h$\alpha$mr). 
We trained the SVM in the multiclass case with a Radial Basis Function (RBF) and implemented using the {\sc scikit-learn} package \citep{scikit-learn11}. We calculated the membership probabilities calibrated using Platt scaling extended for multi-class classification \citep{Wu04}. We transfer the classification probability to the open cluster population. We obtained a final sample of 168 open clusters with high-probability of belonging to the thin disc (P > 0.9) in the metallicity range [Fe/H]=[$-$0.74, 0.45]. Their location in the Tinsley-Wallerstein diagram (TWD) is shown in Fig.~\ref{fig:alphafe_ocs_field_thin}, together with $\sim$200,000 main sequence turn off (MSTO) field stars potentially from the thin disc (see Section~\ref{sec_f_stars}). Of these, 138 OCs (82\%) are located in the Galactocentric interval 6 kpc < R$_{\rm GC}$ < 11 kpc. 

\begin{figure}
  \resizebox{\hsize}{!}{\includegraphics{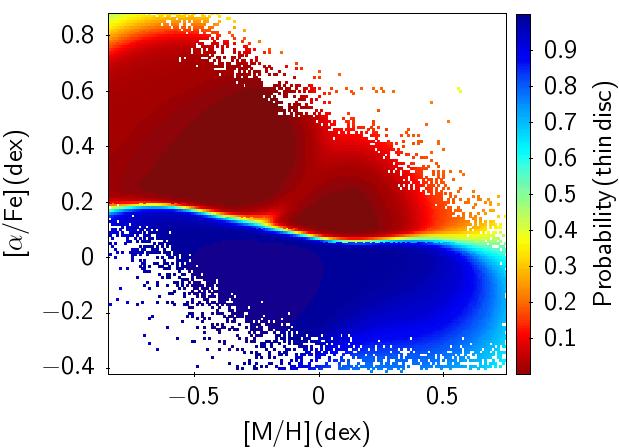}}
  \caption{$\sim$ 900,000 MSTO field stars in the [$\alpha$/Fe] vs [M/H] plane, color coded by the probability of belonging to the thin disc.}
  \label{fig:probability_thin}
\end{figure}

A similar analysis can be done using [Ca/Fe] instead of [$\alpha$/Fe] since the two ratios are very close 
(CaII IR triplet is the dominant source of $\alpha$-element abundances in the {\em Gaia} spectral range). 
In Van der Swaelmen et al. (2023), we preformed  the thin-thick disc separation using both [Ca/Fe] and [$\alpha$/Fe] finding very similar results for the open cluster population. The result is actually expected, since as mentioned above, in {\em Gaia} Ca abundance is the dominant contributor to [$\alpha$/Fe]. 
On the other hand, the use of other $\alpha$ elements might give slightly different results, such as Mg, which shows a different growth than the other $\alpha$ elements at super-solar metallicities \citep[see, e.g.][]{magrini17, palla22}. 
Finally, we might mention that we did not include the treatment of the uncertainties in applying the SVMs analysis to separate the two disc populations since we are interested only in a statistical separation and it would make the analysis more difficult. The choice of a membership probability P > 0.9 implies that stars or clusters at the edge between the two populations are automatically excluded, and only those with a high probability of belonging to one disc or the other are considered (see Figure \ref{fig:probability_thin}). This especially happens for field stars, as the majority of the clusters  have a high probability of belonging to the thin disc. 
The properties of the final sample of clusters used throughout this study are included in the table \ref{tab:OCproperties} in the Appendix.

\subsection{The sample of field stars}
\label{sec_f_stars}

\begin{figure}
  \resizebox{\hsize}{!}{\includegraphics{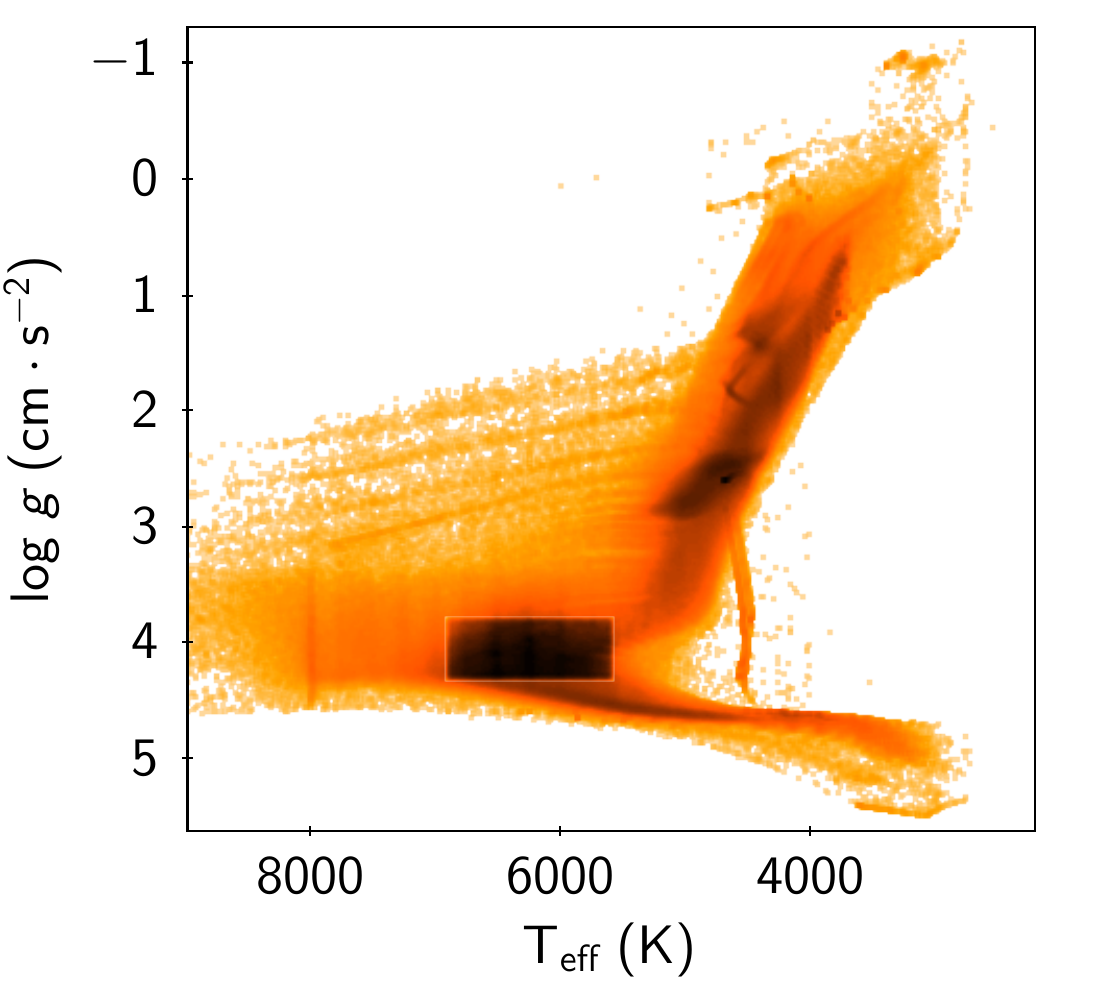}}
  \caption{KD for the more than 5 million field stars of {\em Gaia} {\sc DR3}. The location of the $\sim$ 900,000 MSTO stars is indicated with a red  box. }
  \label{fig:5_mill_KD}
\end{figure}

We select nearly 5 million stars included in the General Stellar Parametrizer (GSP)-Spec sample \citep{Recio-Blanco_2016} belonging to the catalogue of about 30 million stars in the Radial Velocity Spectrometer (RVS).  Among them, we select those located around the MSTO, as shown in Fig.~\ref{fig:5_mill_KD}, since their ages are expected to be more accurate and reliable than those of stars in different evolutionary phases \citep[e.g.][]{Howes_2019}. To perform this selection, we consider stars that have log$g$ between 3.8 and 4.3, and  $T_{\mathrm{eff}}$ between 5600 and 6900 K as in \citet{Chen_2022}. This reduced the sample to about 900,000 stars (16 \% of the total sample). Fig.~\ref{fig:5_mill_KD} shows the Kiel Diagram (KD) for the $\sim$5 million field stars with the $\sim$900,000 MSTOs stars boxed. Of these $\sim$900,000 stars we selected $\sim$200,000 potentially belonging to the thin disc (P>0.9), using the same techniques and the same training set as for open clusters. In this way, the field stars sample results in a very similar metallicity range ([-0.86, 0.59]) to the open clusters' one. About 99\% of the stars in the selected MSTO-thin disc sample are located between 7.5 kpc < R$_{\rm GC}$ < 9 kpc (expressed in the catalogue by the column R\_med\_dgeo). 
From them, we extracted only stars whose ages were determined in \citet{Kordopatis_2023} (see Sec.~\ref{sec_age} for the age determination and selection) and we applied to them the HQ and MQ selection defined in \citet{Gaia23_chemical_cartography}. The final sample consists of $\sim$66,000 stars, selected in terms of HQ and MQ in the same way as the  open cluster member stars. 
Finally,  we ensured that we have consistent samples in terms of positions in the Galaxy with 'a posteriori' selection, i.e. after applying the quality selections, we reduced the sample of the field stars to the thin disc and the galactocentric region that we also want to map with clusters. 

\subsubsection{Ages of field stars}
\label{sec_age}
For a high-confidence age determination, we selected a sub-sample of stars that meets the following characteristics: the standard deviation, $\sigma$, of the five ages calculated in \citet{Kordopatis_2023} considering different types of projection (with different combinations of absolute magnitudes JHKG) and the age of the stars from the Final Luminosity Age Mass Estimator (FLAME) \citep{Andrae_2018} provided by \texttt{gaiadr3.astrophysical\_parameters} should be less than 1 Gyr. 
We selected the average of the six determinations as our final age, and we used  the standard deviation as its uncertainty. In addition, we consider, as for open clusters, only stars with ages > 1~Gyr.
It is important to note that the ages published in \citet{Kordopatis_2023} are obtained from calibrated stellar parameters, while the FLAME ages do not.  This may have a non-negligible effect for the ages of giants, for which the effect of calibrated log$g$ is larger, but should be minimal in the case of MSTOs. 
By construction, our sample of field stars contains stars with uncertainties in age less than 1 Gyr. These values are comparable to those obtained for clusters. From the paper of \citet{CantatGaudin20}, the uncertainty on the determination of $\log$(age) ranges from 0.15 to 0.25 for young clusters and from 0.1 to 0.2 for old clusters. Considering that in our sample we have only ‘old’ clusters, with age > 1 Gyr, the typical uncertainties range from 0.2-0.3 Gyr for the youngest ones in our samples to about 1 Gyr for the oldest ones.  They are, therefore, comparable and consistent with the uncertainties of the selected field star sample.

\begin{figure}
  \resizebox{\hsize}{!}{\includegraphics{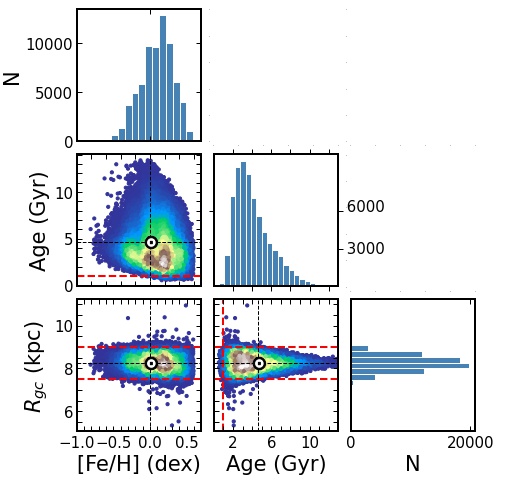}}
  \caption{Relationships between the properties of the $\sim$70,000 selected field stars: number of stars, ages and Galactocentric distances. The dashed red lines indicate the cutoff in age and R$_{\rm GC}$.}
  \label{fig:69617_pairwisse}
\end{figure}

Fig.~\ref{fig:69617_pairwisse} shows the relationships between age, Galactocentric distance R$_{\rm GC}$, and metallicity for the selected samples. Both clusters and field stars occupy a similar range of metallicity. However, the age of field stars span a wider range
than for clusters, since clusters generally do not survive beyond 7 Gyr. On the other hand, the selected clusters sample a wider range of
Galactocentric distances since the cluster member stars are mostly luminous giants, i.e. they are easier to observe at large distances than the selected MSTO field stars. 
Finally, the difference in the types of stars observed in the field and in the clusters could generate observational biases, for example, in the derived metallicities. 
In parallel to the present paper, we are carrying out a work aiming to compare and validate {\em Gaia} spectroscopic parameters by comparing them with the  {\em Gaia}-ESO ones (Van der Swaelmen et al. 2023). 
We confirm that there is an excellent agreement between the calibrated spectroscopic metallicities and [$\alpha$/Fe] of both giants and dwarfs in {\em Gaia}-ESO and in {\em Gaia}. So, taking the  {\em Gaia}-ESO survey as a reference, there are no systematic differences in  {\em Gaia}'s calibrated metallicities and [$\alpha$/Fe] considering giants and dwarfs.

\section{Comparing the kinematic properties of clusters and field stars}
\label{section:properties}

In this section, we compare the evolution over time of the space velocities, orbital parameters, and orbital action of a reduced sample of clusters  and of sample of field stars located in the same Galactocentric region. 

The radial distribution in the Qxx  quantiles (percentiles) of the  $\sim$70,000 field stars is R$_{\rm GC}$ [Q01, Q10, Q90, Q99]    [7.53, 7.82, 8.69, 8.96](kpc). Thus, 98\% of the field stars ($\sim$68,000) are located between 7.53 and 8.96 kpc, from which we select 66,000 HQ and MQ stars. We selected the corresponding  41 clusters (see Fig.~\ref{fig:alphafe_ocs_field_thin}) located in the regions [7.5-9]~kpc.

\begin{figure}
  \resizebox{\hsize}{!}{\includegraphics{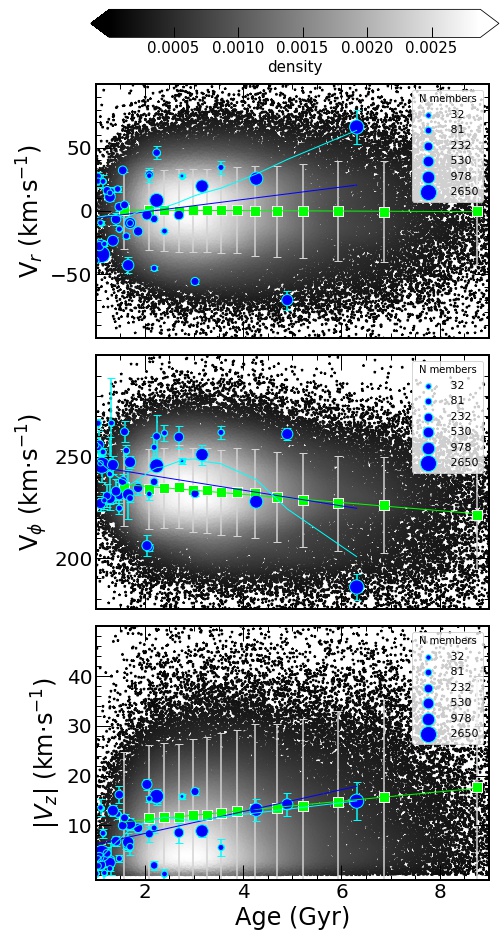}}
  \caption{Space velocities (V$_R$, V$_{\phi}$ and |V$_Z$|) for $\sim$66,000 field stars from our selected sample and 41 open clusters in the solar region. The data is presented in equally distributed bins (q=14) for field stars (lime). In the background the field stars are also shown on a density plot which is encoded in the colourbar. The size of the symbols for clusters (blue circles) are  proportional to the square root of their number of members ($\sqrt{N}$), shown in the legend with their total number of members. The straight lines represent the linear fits (green for field stars and blue for open clusters) and the curve (cyan) is a nonparametric lowess model (locally weighted linear regression) to the clusters' data}.
  \label{fig:OCs_vs_field_V_vs_age_solar}
\end{figure}

\begin{figure}
  \resizebox{\hsize}{!}{\includegraphics{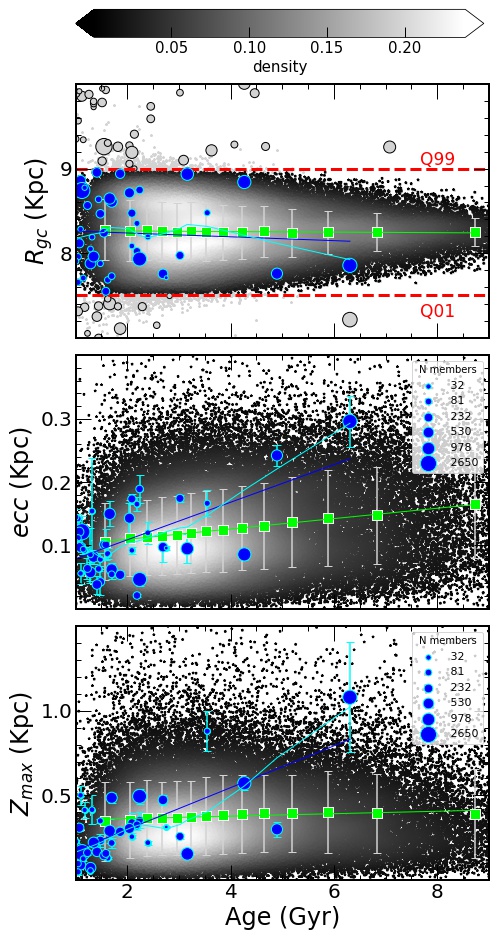}}
  \caption{Orbital parameters ($R$, $e$ and $Z_{max}$) for $\sim$66,000 field stars from our selected sample and $\sim$40 open clusters in the solar region. The data are presented in equally distributed bins (q=14) for field stars (lime squares). In the background the field stars are also shown on a density plot which is encoded in the colourbar. Symbols and colours are as in Fig.~\ref{fig:OCs_vs_field_V_vs_age_solar}.} 
  \label{fig:OCs_vs_field_R_e_zmax_vs_age_solar}
\end{figure}

\begin{figure}
  \resizebox{\hsize}{!}{\includegraphics{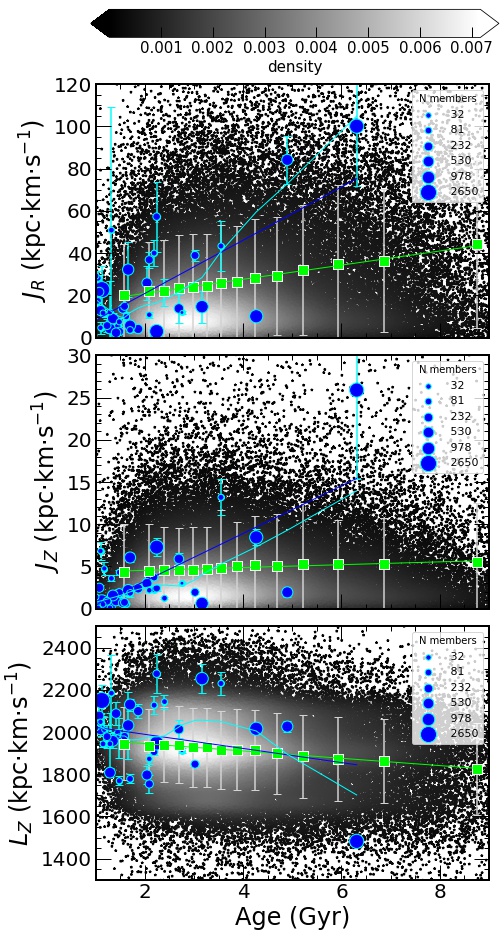}}
  \caption{Orbital actions ($J_{R}$, $J_{Z}$ and $L_{Z}$) for $\sim$66,000 field stars from our selected sample and $\sim$40 open clusters in the solar region. The data is presented in equally distributed bins (q=14) for field stars (lime). In the background the field stars are also shown on a density plot which is encoded in the colourbar.  Symbols and colours are as in Fig.~\ref{fig:OCs_vs_field_V_vs_age_solar}.} 
  \label{fig:OCs_vs_field_Jr_vs_age_solar}
\end{figure}

The three-dimensional galactocentric coordinates, (cylindrical) space velocities, and orbital parameters of the sample field stars are obtained from  \citet{Gaia23_chemical_cartography}. 
The Galactocentric coordinates X, Y, Z (in Cartesian coordinates), the Galactocentric distance (R$_{\rm GC}$), and cylindrical space velocities (radial V$_R$, azimuthal V$_{\phi}$, vertical V$_Z$) in a right-handed are computed from the right ascension, declination, line-of-sight velocity, proper motions, and the EGDR3 geometric and photogeometric Bayesian line-of-sight distances from \citet{bailerjones21}.  
They assumed a solar position (R, Z)$_{\odot}$=(8.249, 0.0208) kpc and the solar cylindrical velocity components are (V$_R$, V$_{\phi}$, V$_Z$)$_{\odot}$=(9.5, 250.7, 8.56) km s$^{-1}$ \citep{gravity20}.
Their orbital parameters were computed with the {\sc Galpy} code (Bovy 2015), using the axisymmetric Galactic potential of \citet{mcmillan17}. 
For the clusters, we compute the orbits in a consistent way with {\sc galpy}, using the clusters mean parallaxes, radial velocities and distances from {\em Gaia}. 
Due to the large number of field stars and their high density, we use a point density function (\texttt{gaussian\_kde}) to represent them (see Figs.  \ref{fig:OCs_vs_field_V_vs_age_solar} to \ref{fig:OCs_vs_field_Jr_vs_age_solar}), implemented using \texttt{scipy.stats}, determining the density of stars at each point and assigning that value to the colours in the colourmap. For an easier comparison, the $\sim$66,000 field stars are also shown in equally distributed bins using a 'Quantile-based discretization function', defining the bins using percentiles based on the distribution of the data. We divided the data into 14 quantiles (q) of approximately $\sim$5,000 stars each and computed the mean and dispersion for each bin. We also show regressions (linear and non-linear) applied to both samples using the Ordinary Least Squares (OLS) method and a nonparametric LOWESS model (locally weighted linear regression) respectively, implemented using \texttt{statsmodels}. For a better comparison we also apply a Pearson and Spearman statistical correlation test computed using \texttt{scipy.stats} to measure the strength and direction of the relationship (linear and monotonic) between variables. In Figures \ref{fig:KS_Age} to \ref{fig:KS_Lz} in the Appendix \ref{Appendix} we also show the cumulative distribution functions (CDFs) and the results of the two-sample Kolmogorov-Smirnov (K-S) test statistic \citep{kolmogorov33, smirnov1939estimate} computed using \texttt{scipy.stats}. This allows us to analyze in more detail the distribution of the data of both samples and find where is the maximum absolute difference between the two cumulative distributions.


\begin{table}
\caption{Linear regression coefficients (slope and y-intercept) obtained using the least squares method, as well as Pearson and Spearman correlation coefficients and their p-values for both clusters and field stars samples.}
\label{tab:regressions}
\scalebox{0.77}{
\begin{tabular}{lcccccc}

\hline\hline
& & & field stars & & & \\
\hline
Param. & m & c & PCC & p-value & SCC & p-value \\
\hline
V$_R$ & $-$0.162 & 0.899 & $-$0.009& 0.020& $-$0.007& 0.067 \\  
V$_{\phi}$ & $-$1.999& 239.419& $-$0.177& 0.000& $-$0.151&0.000  \\ 
|V$_Z$| & +0.947& 9.162&+0.176 & 0.000&+0.149 &  0.000\\ 
$R$ &$-$0.003 & 8.270&$-$0.021 &0.000 &$-$0.025 & 0.000 \\ 
$e$ & +0.008& 0.094&+0.227 & 0.000& +0.200& 0.000 \\
$Z_{max}$ &+0.007 & 0.350& +0.058& 0.000& +0.069& 0.000 \\ 
$J_{R}$ & +3.330& 14.316& +0.220& 0.000& +0.196& 0.000 \\
$J_{Z}$ & +0.166& 4.213& +0.056& 0.000& +0.067& 0.000 \\
$L_{Z}$ & $-$17.197& 1979.141& $-$0.175& 0.000& $-$0.154& 0.000 \\ 

\hline
& & & open clusters & & & \\
\hline
Param. & m & c & PCC & p-value & SCC & p-value \\
\hline
V$_R$ & +4.910& $-$10.597& +0.189& 0.237&+0.192 & 0.229 \\ 
V$_{\phi}$ & $-$3.882& +249.036& $-$0.247& 0.120& $-$0.034& 0.831 \\ 
|V$_Z$| & +2.203&+3.931 & +0.455& 0.003& +0.530& 0.000 \\ 
$R$ & $-$0.023& 8.284& $-$0.059&0.713 & +0.023& 0.887 \\ 
$e$ & +0.030& 0.049& +0.594&0.000 &+0.224 &0.158  \\
$Z_{max}$ & +0.124& 0.051& +0.677& 0.000& +0.508& 0.001 \\ 
$J_{R}$ & +12.608& $-$4.284 &+0.654 &0.000 &+0.211 & 0.185 \\ 
$J_{Z}$ & +2.821& $-$2.375&+0.703 &0.000 & +0.505& 0.001 \\ 
$L_{Z}$ & $-$33.421& + 2055.896& $-$0.248& 0.118&$-$0.079 & 0.623 \\

\hline
\end{tabular}}
\end{table}

\subsection{Space Velocities over time}

Stars and open clusters orbit the Galactic Center in quasi circular orbits. The distribution of velocities of stars in the thin disc of the Galaxy is a function of the age of the stars. Younger stars, which formed relatively recently, have tangential velocities that are close to those the Galactic disc at solar location and the Sun \citep[i.e., $\sim$240-250 km~s$^{-1}$][]{russel17, gravity20}. This is because they are on less perturbed orbits. Instead, older stars typically have smaller tangential velocities  because they are likely on more elliptical orbits. Here we aim at comparing the velocity components of stars and clusters to investigate if there are substantial differences in their behaviour. 
In Fig.~\ref{fig:OCs_vs_field_V_vs_age_solar} we show the three components of the Galactic space velocities of star clusters compared to the ones of field stars, as a function of stellar ages. 
In Table \ref{tab:regressions} we show the coefficients of the linear regressions and the Pearson and Spearman correlation coefficient (PCC and SCC, respectively) with associated p-value.
We recall that with a p-value  less than 0.05, the results are considered to be statistically significant, implying that the observed correlation is not simply the result of chance and is more likely to reflect a real relationship between the variables. If the p-value is greater than 0.05, the results are not statistically significant, which means that there is not enough evidence to reject the null hypothesis and it cannot be affirmed that there is a significant relationship between the variables.

Field stars and open clusters show correlations 
in their trends of radial V$_R$, tangential V$_{\phi}$ and vertical V$_Z$ velocities  as a function of stellar ages, which are in some cases
different (see the coefficients of the linear regressions in Table~\ref{tab:regressions}). Some of these differences also have statistical evidence (low p-values), while others are statistically weaker, likely due to the limited number of old clusters and to the scatter of their properties.

The average radial velocity (upper panel) of field stars is quite constant over time, with a scatter increasing in the older populations. Young clusters have a slightly lower radial component than field stars, and show significant scatter. Both correlations are not very strong or statistically significant in both cases (high p-values in Table~\ref{tab:regressions}).
The tangential velocity component, V$_{\phi}$ (central panel) is close to that of the Milky Way disc rotation for objects moving along circular orbits at the Sun location \citep{russel17}. Several young clusters  have  higher V$_{\phi}$ than the one of field stars in the same age range, and the intercept of the regression for star clusters in higher than that of field stars (249 km~$s^{-1}$ vs 239 km~$s^{-1}$). 
As the ages of clusters and stars increase, the two trends tend to converge, although the number of old clusters decreases dramatically after 3 Gyr. This is well reflected in the LOWESS model which is able to keep the changes between the behaviour of young and old clusters. In this case,  the linear  correlation is statistically significant for field stars (with very low p-values), while it is not for clusters. 
Concerning the absolute vertical velocity, |V$_Z$| (bottom panel), it increase in both populations over time.
However, young clusters have a smaller vertical velocity component than field stars, reaching them only for ages above 4 Gyr. In this case, the linear correlations are statistically significant in both samples. The clusters show higher PCC and SCC than the field stars, indicating a stronger correlation (see Table \ref{tab:regressions}).
The combination of the three results, obviously interconnected, shows us that during the first 1-3 Gyr star clusters remain more stably in nearly circular orbits than single stars, while older clusters have typically more perturbed velocity components than field stars. 


\subsection{Orbital parameters and actions over time}
\label{subsection:orbital_param}

Using the velocity components and the distance of a star, and assuming  a gravitational potential, the orbit of a star, characterised by its guiding radius $R$, its eccentricity $e$ and its inclination, parameterised by the maximum height reached above the Plane, $Z_{max}$, can be derived. Circular orbits have eccentricities closer to zero, and will reach low heights above the Plane. 

In Fig.~\ref{fig:OCs_vs_field_R_e_zmax_vs_age_solar}, we show the Galactocentric radius and the orbital parameters $e$ and $Z_{max}$ as a function of stellar ages. 
In the upper panel, we present  the distribution of R$_{\rm GC}$ as a function of stellar ages. As per sample selection, field stars and clusters are confined between 7.5 and 9~kpc. 
In the central  panel, we show the relation between the eccentricity of the orbit and stellar ages. During the first 3 Gyr, clusters and field stars show a slightly different behaviour: on average the orbits of clusters have lower eccentricities ($e$ < 0.1, on average), i.e. more circular and less perturbed, than those of field stars. However, as time passes, clusters proceed faster towards more eccentric orbits ($e$ > 0.1), or, equivalently, open clusters with low eccentricity do not exist anymore (at least in our sample limited to the Solar neighborhood).
 The correlations are significant in both cases, but the effect is more pronounced in clusters (higher PCC and SCC).
 Finally, in the bottom panel, we show the maximum height, Z$_{max}$,  above the Plane as a function stellar ages. 
 As for the eccentricity of the orbits, younger clusters are orbiting closer to the Galactic Plane than field stars having the same age. However, the situation changes for the surviving clusters beyond $\sim$3 Gyr that reach higher heights above the Galactic Plane, while field stars experience similar but smoother changes over time. The correlations are statistically significant in both cases, but again, the increase is steeper in clusters (higher PCC and SCC).

An equivalent way to the use of the orbital parameters is to describe the motion of a star or of a stellar cluster by their orbital actions, which are three fundamental quantities used to describe the motion of a particle (both star or cluster) in a rotating galaxy. The radial action ($J_{R}$) describes the component of a star's angular momentum in the direction of the Galactic Center, the vertical action ($J_{z}$)  describes the component of a star's angular momentum perpendicular to the Galactic plane, and the azimuthal action ($L_{Z}$, equivalent to $J_{\phi}$)  describes the component of a star's angular momentum around the Galactic Center.
In axisymmetric potentials, the orbital actions are used to quantify the amount of oscillation of the star along its orbit in the Galactocentric  directions \citep[R, $\phi$, z, see][]{binneytremaine08}.
For the interpretation of the orbital actions, we follow \citet{trick19}: 
the radial action $J_{R}$  can be considered as a measure of the orbit eccentricity or the radial extent of a disc orbit’s in-plane epicyclic rosette;  the azimuthal action, $J_{\phi}$, is equivalent to the angular momentum in the z-direction, $L_z$, and it describes the amount of rotation around the Galactic Center. 
Finally, $J_{z}$, the vertical action, quantifies the displacement above and below the Galactic Plane. 

In Fig.~\ref{fig:OCs_vs_field_Jr_vs_age_solar} we show the orbital actions of the clusters compared to the field stars. 
In the upper panel, we show the radial action $J_{R}$ over time. As said above, its behaviour is similar to that of the eccentricity: young clusters have typically lower $J_{R}$ than field stars. The correlations are statistically significant in both cases, with a steeper growth  for  clusters (higher PCC and SCC). 
In the central panel, we present  the vertical action $J_{z}$, that indicates the displacement above and below the Galactic Plane. 
Also in this case, the younger clusters of our sample do not exhibit a large vertical excursion around the Plane, while the trend indicates that older clusters are more likely to explore regions far from the Plane due to their inclined orbit. As in the |V$_Z$| case, the correlations are statistically significant in both cases (p-values < 0.05), but stronger for the clusters case (higher PCC and SCC).
Finally, $L_{Z}$ in young clusters is larger than in field stars thus again indicating that the orbit of clusters is closer to a circular orbit than that of field stars. However, due to the large scatter in the cluster data, the relationship between $L_{Z}$ and age have very low statistical relevance, confirmed by the high p-values (0.118 for PCC and 0.623 for SCC). 

\section{Discussion on the old survived clusters}
\label{section:discussion}

There is some statistical evidence that correlations exist between kinematic properties of clusters and field stars and their ages, and that, in some cases,  such correlation might differ indicating a different behaviour of field stars and clusters. In other cases, these differences do not have sufficient statistical value.
From the comparison of the kinematic and orbital properties of the cluster and field star population (in particular V$_Z$, Z$_{\rm max}$ and J$_Z$, for which PCC and SCC have low p-values), we can conclude that the former are on average more resistant to perturbative effects up to an age of about 3 Gyr, moving on  quasi-circular orbits, close to the Galactic Plane. On the other hand, clusters older than 3~Gyr are quite scattered on the kinematical properties-age planes, with some of them having orbits with higher eccentricity, more inclined, thus reaching higher heights on the Plane. The fact that several old clusters have  eccentric orbits ($e>$0.15) is not a cause in itself, but rather it is likely a consequence of the passage of time, and a necessary condition to allow their survival \citep{Maxwell_2016}.  
Several authors claimed that even in the first Myr interaction with molecular clouds are  more disruptive for low-mass clusters \citep[see, e.g.][]{gieles16}, and only massive clusters with  peculiar orbits might survive to the several interactions that happen in the Galactic disc in the following Gyr \citep{moitinho10, Buckner14}. The reason why they now stand out might be related to a natural selection effect, since clusters  located closer to the Galactic Plane would have more interaction and thus dissolve more rapidly. But why don't interactions, which are proven to cause such drastic
changes in the orbit, lead to the destruction of the cluster  as \citet{friel95} pointed out? What are the particular physical properties that have made these
clusters survive until today? 
\citet{gustafsson16} demonstrated that just a small fraction of massive clusters can survive for several Gyr, and that only 0.5\% of all formed massive open clusters are predicted to end with high altitude on the Plane. 

\begin{figure*}
  \resizebox{\hsize}{!}{\includegraphics{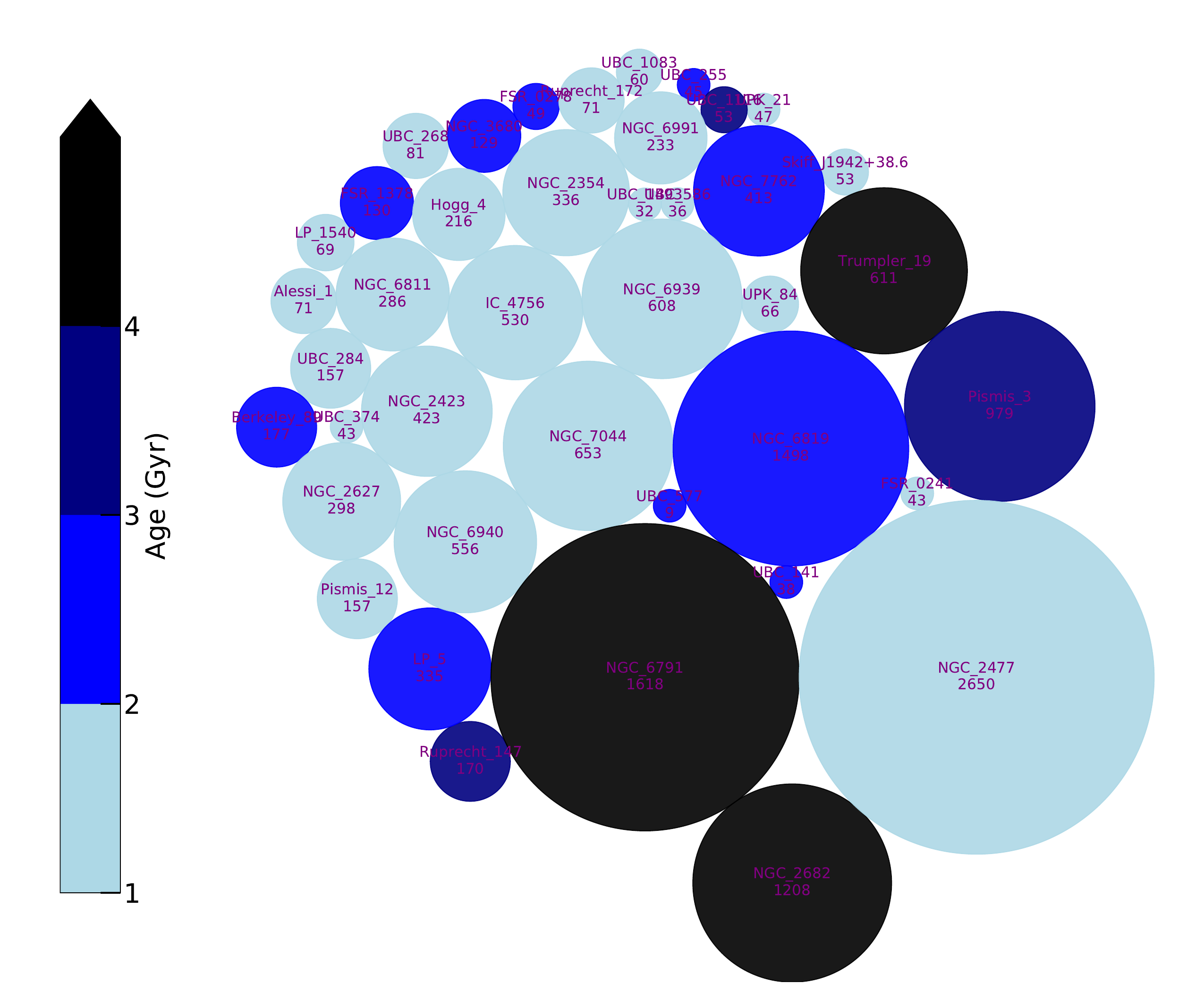}}
  \caption{Our selected sample of 41 open clusters on the thin disc and solar region sized by the total number of members, coloured by age, and labelled by cluster name and number of member stars considered in this study.}
  \label{fig:clusters_by_N_members_age_solar}
\end{figure*}

To seek answer to these questions, we can examine some of the examples of the surviving old clusters to get an idea of their general characteristics. Old open clusters are, indeed,  rare  as star clusters dissipate over time. We expect that  only the most massive, dense, and well-placed ones can survive several Gyr \citep{Boesgaard_2015}. In Fig.~\ref{fig:clusters_by_N_members_age_solar},  we plot our  sample of 41 clusters with symbols proportional to their number of members (as estimated from {\em Gaia} in \citet{CantatGaudin20}).
Most clusters older than 3 Gyr have a high average number of members. 
Younger clusters have a more variable number of members, ranging from the highly populated clusters, such as  NGC\,2477, to clusters with few members such as UBC\,139 \citep[members from][]{CantatGaudin20}. 


Among the oldest clusters, there are some clusters that stand out because they are more populous that the other clusters: NGC\,6791, NGC\,2682 (M67), and Trumpler~19. Their large number of members has been confirmed also using methods based on the {\sc DBSCAN} algorithm, complementary to the kinematic methods \citep{Gao_2014}. 
The common characteristics of  these clusters are that they are currently high above the Galactic Plane, still have a high density and a large number of members, and a relatively high metallicity. 
In particular, NGC 2682 is located in a low density region, and it has not likely experienced significant gravitational interactions that could have affected its structure \citep[cf.][]{Davenport_2010}. Recent works revealed that it is more massive than previously believed \citep{Carrera_2019} and it underwent a mass segregation process \citep[e.g.][]{ Geller_2015}, which on long term timescale could make the cluster tightly bound and less likely to disperse.
Finally, the oldest and highly populated cluster in our sample is NGC~6791, which  is  among  the most studied clusters due to its various peculiarities, as the high eccentricity and maximum height above the Plane, coupled with a high metallicity. Indeed, NGC~6791 has an orbit similar to that of a globular cluster or a dwarf galaxy than to that of a thin disc open cluster \citep{Carraro_1994, Jílkova_2012}. \citet{gustafsson16} suggested that NGC~6791  maintained so many members because several generations of stars have formed within it, given that the material expelled by AGB stars inside the cluster was retained within it. This would later be the seed for the birth of new generations of stars. However, there are no evidence of chemical anomalies or abundance large spread in NGC~6791 \citep{carretta07, bragaglia14}.
As seen in \citet{Viscasillas_2022}, NGC~6791 is chemically misplaced, not only for its higher metallicity, but also for its [$\alpha$/{\em slow-neutron capture}] element ratios, which does not agree to that of clusters found in the same region, hence emphasises the importance of taking migration into account in chemical evolution studies.
So, in conclusion, it appears that in the solar neighbourhood the oldest clusters that managed to survive have in common a large initial mass, and fortuitous orbital conditions \citep[cf.][]{van_den_Bergh_1980}.
However, in different parts of the Galaxy, like the outer Galaxy, our ability to observe clusters is linked to possible observational biases, which favours the detection of massive, distant clusters high on the Plane.





\section{Summary and Conclusions}
\label{sec:conclusions}
We conducted a purely observational study using high-quality spectroscopic {\em Gaia} {\sc DR3} data to identify the differences between the  kinematics of the population of  field stars and open clusters. 
For a meaningful comparison between the kinematic and dynamical properties of clusters and field stars, we restricted our sample to the radial region [7.5--9~kpc]. 
Furthermore, we restricted the sample to clusters older than 1 Gyr because our aim is to estimate the effect of migration, which is negligible for younger clusters. 
We selected the sample of field stars around the MSTO, so as to have a better determination of their ages. 
We compared the velocity components, the orbital parameters and orbital actions of our sample of $\sim$66,000 field stars and of 41 open clusters. We conducted a statistical analysis to test the significance of the results.
Open star clusters younger than 2-3 Gyr maintain  circular orbits, with a dominant tangential component V$_{\phi}$ of their velocity than in field stars with similar ages. This corresponds to more circular orbits (lower eccentricity and lower height from the Galactic Plane), and it is also reflected on the orbital actions. In particular, we observed lower $J_R$ and $J_Z$ for the clusters with ages below 2-3 Gyr than for field stars of the same age. 
On the other hand, older clusters are less dominant by number and are characterised by more perturbed orbits and to be found typically higher on the Plane. 
These characteristics, together with being massive --or having been massive--, seem to be essential to ensure their survival for several Gyr. 
Thus, the oldest clusters, although still chemical tracers of the Galaxy's past composition, do not reflect the composition of the place where they are currently found. 
As already noticed in \citet{magrini23}, the radial metallicity gradient of clusters older than 3 Gyr shows a higher level of scatter, and it is not obvious to use it to study the temporal evolution of the gradient unless kinematic constraints are also considered.

\begin{acknowledgements}
The authors thank the anonymous referee for her/his constructive and insightful suggestions which greatly improved the paper.
This work has made use of data from the European Space Agency (ESA) mission
{\it Gaia} (\url{https://www.cosmos.esa.int/gaia}), processed by the {\it Gaia} Data Processing and Analysis Consortium (DPAC, \url{https://www.cosmos.esa.int/web/gaia/dpac/consortium}).Funding for the DPAC has been provided by national institutions, in particular the institutions
participating in the {\it Gaia} Multilateral Agreement.
This work also made use of SciPy \citep{SciPy20}, Astropy \citep{Astropy_2018}, Scikit-learn Machine Learning \citep{scikit-learn11}, StatsModels \citep{seabold2010statsmodels}, Seaborn \citep{Waskom2021}, TopCat \citep{Taylor2005}, Pandas \citep{pandas2020} and Matplotlib \citep{Hunter2007}.  CVV and LS thanks the EU programme Erasmus+ Staff Mobility for their support. CVV and GT acknowledge funding from the Lithuanian Science Council (LMTLT, grant No. P-MIP-23-24).
LM thanks INAF for the support (MiniGrant Checs). 

\end{acknowledgements}

\bibliographystyle{aa} 
\bibliography{Bibliography}

\begin{thebibliography}{81}
\expandafter\ifx\csname natexlab\endcsname\relax\def\natexlab#1{#1}\fi

\bibitem[{{Ahn} {et~al.}(2014){Ahn}, {Alexandroff}, {Allende Prieto}, {Anders},
  {Anderson}, {Anderton}, {Andrews}, {Aubourg}, {Bailey}, {Bastien},
  {Bautista}, {Beers}, {Beifiori}, {Bender}, {Berlind}, {Beutler}, {Bhardwaj},
  {Bird}, {Bizyaev}, {Blake}, {Blanton}, {Blomqvist}, {Bochanski}, {Bolton},
  {Borde}, {Bovy}, {Shelden Bradley}, {Brandt}, {Brauer}, {Brinkmann},
  {Brownstein}, {Busca}, {Carithers}, {Carlberg}, {Carnero}, {Carr},
  {Chiappini}, {Chojnowski}, {Chuang}, {Comparat}, {Crepp}, {Cristiani},
  {Croft}, {Cuesta}, {Cunha}, {da Costa}, {Dawson}, {De Lee}, {Dean},
  {Delubac}, {Deshpande}, {Dhital}, {Ealet}, {Ebelke}, {Edmondson},
  {Eisenstein}, {Epstein}, {Escoffier}, {Esposito}, {Evans}, {Fabbian}, {Fan},
  {Favole}, {Femen{\'\i}a Castell{\'a}}, {Fern{\'a}ndez Alvar}, {Feuillet},
  {Filiz Ak}, {Finley}, {Fleming}, {Font-Ribera}, {Frinchaboy},
  {Galbraith-Frew}, {Garc{\'\i}a-Hern{\'a}ndez}, {Garc{\'\i}a P{\'e}rez}, {Ge},
  {G{\'e}nova-Santos}, {Gillespie}, {Girardi}, {Gonz{\'a}lez Hern{\'a}ndez},
  {Gott}, {Gunn}, {Guo}, {Halverson}, {Harding}, {Harris}, {Hasselquist},
  {Hawley}, {Hayden}, {Hearty}, {Herrero Dav{\'o}}, {Ho}, {Hogg}, {Holtzman},
  {Honscheid}, {Huehnerhoff}, {Ivans}, {Jackson}, {Jiang}, {Johnson},
  {Kinemuchi}, {Kirkby}, {Klaene}, {Kneib}, {Koesterke}, {Lan}, {Lang}, {Le
  Goff}, {Leauthaud}, {Lee}, {Lee}, {Long}, {Loomis}, {Lucatello}, {Lupton},
  {Ma}, {Mack}, {Mahadevan}, {Maia}, {Majewski}, {Malanushenko},
  {Malanushenko}, {Manchado}, {Manera}, {Maraston}, {Margala}, {Martell},
  {Masters}, {McBride}, {McGreer}, {McMahon}, {M{\'e}nard}, {M{\'e}sz{\'a}ros},
  {Miralda-Escud{\'e}}, {Miyatake}, {Montero-Dorta}, {Montesano}, {More},
  {Morrison}, {Muna}, {Munn}, {Myers}, {Nguyen}, {Nichol}, {Nidever},
  {Noterdaeme}, {Nuza}, {O'Connell}, {O'Connell}, {O'Connell}, {Olmstead},
  {Oravetz}, {Owen}, {Padmanabhan}, {Palanque-Delabrouille}, {Pan}, {Parejko},
  {Parihar}, {P{\^a}ris}, {Pepper}, {Percival}, {P{\'e}rez-R{\`a}fols}, {Dotto
  Perottoni}, {Petitjean}, {Pieri}, {Pinsonneault}, {Prada}, {Price-Whelan},
  {Raddick}, {Rahman}, {Rebolo}, {Reid}, {Richards}, {Riffel}, {Robin},
  {Rocha-Pinto}, {Rockosi}, {Roe}, {Ross}, {Ross}, {Rossi}, {Roy},
  {Rubi{\~n}o-Martin}, {Sabiu}, {S{\'a}nchez}, {Santiago}, {Sayres},
  {Schiavon}, {Schlegel}, {Schlesinger}, {Schmidt}, {Schneider}, {Schultheis},
  {Sellgren}, {Seo}, {Shen}, {Shetrone}, {Shu}, {Simmons}, {Skrutskie},
  {Slosar}, {Smith}, {Snedden}, {Sobeck}, {Sobreira}, {Stassun}, {Steinmetz},
  {Strauss}, {Streblyanska}, {Suzuki}, {Swanson}, {Terrien}, {Thakar},
  {Thomas}, {Thompson}, {Tinker}, {Tojeiro}, {Troup}, {Vandenberg}, {Vargas
  Maga{\~n}a}, {Viel}, {Vogt}, {Wake}, {Weaver}, {Weinberg}, {Weiner}, {White},
  {White}, {Wilson}, {Wisniewski}, {Wood-Vasey}, {Y{\`e}che}, {York}, {Zamora},
  {Zasowski}, {Zehavi}, {Zhao}, {Zheng}, \& {Zhu}}]{Ahn_2014}
{Ahn}, C.~P., {Alexandroff}, R., {Allende Prieto}, C., {et~al.} 2014, \apjs,
  211, 17

\bibitem[{{Allison} {et~al.}(2010){Allison}, {Goodwin}, {Parker}, {Portegies
  Zwart}, \& {de Grijs}}]{Allison_2010}
{Allison}, R.~J., {Goodwin}, S.~P., {Parker}, R.~J., {Portegies Zwart}, S.~F.,
  \& {de Grijs}, R. 2010, \mnras, 407, 1098

\bibitem[{{Andrae} {et~al.}(2018){Andrae}, {Fouesneau}, {Creevey}, {Ordenovic},
  {Mary}, {Burlacu}, {Chaoul}, {Jean-Antoine-Piccolo}, {Kordopatis}, {Korn},
  {Lebreton}, {Panem}, {Pichon}, {Th{\'e}venin}, {Walmsley}, \&
  {Bailer-Jones}}]{Andrae_2018}
{Andrae}, R., {Fouesneau}, M., {Creevey}, O., {et~al.} 2018, \aap, 616, A8

\bibitem[{{Angelo} {et~al.}(2023){Angelo}, {Santos}, {Maia}, \&
  {Corradi}}]{angelos23}
{Angelo}, M.~S., {Santos}, J.~F.~C., J., {Maia}, F.~F.~S., \& {Corradi},
  W.~J.~B. 2023, \mnras, 522, 956

\bibitem[{{Astropy Collaboration} {et~al.}(2018){Astropy Collaboration},
  {Price-Whelan}, {Sip{\H{o}}cz}, {G{\"u}nther}, {Lim}, {Crawford}, {Conseil},
  {Shupe}, {Craig}, {Dencheva}, {Ginsburg}, {VanderPlas}, {Bradley},
  {P{\'e}rez-Su{\'a}rez}, {de Val-Borro}, {Aldcroft}, {Cruz}, {Robitaille},
  {Tollerud}, {Ardelean}, {Babej}, {Bach}, {Bachetti}, {Bakanov}, {Bamford},
  {Barentsen}, {Barmby}, {Baumbach}, {Berry}, {Biscani}, {Boquien}, {Bostroem},
  {Bouma}, {Brammer}, {Bray}, {Breytenbach}, {Buddelmeijer}, {Burke},
  {Calderone}, {Cano Rodr{\'\i}guez}, {Cara}, {Cardoso}, {Cheedella}, {Copin},
  {Corrales}, {Crichton}, {D'Avella}, {Deil}, {Depagne}, {Dietrich}, {Donath},
  {Droettboom}, {Earl}, {Erben}, {Fabbro}, {Ferreira}, {Finethy}, {Fox},
  {Garrison}, {Gibbons}, {Goldstein}, {Gommers}, {Greco}, {Greenfield},
  {Groener}, {Grollier}, {Hagen}, {Hirst}, {Homeier}, {Horton}, {Hosseinzadeh},
  {Hu}, {Hunkeler}, {Ivezi{\'c}}, {Jain}, {Jenness}, {Kanarek}, {Kendrew},
  {Kern}, {Kerzendorf}, {Khvalko}, {King}, {Kirkby}, {Kulkarni}, {Kumar},
  {Lee}, {Lenz}, {Littlefair}, {Ma}, {Macleod}, {Mastropietro}, {McCully},
  {Montagnac}, {Morris}, {Mueller}, {Mumford}, {Muna}, {Murphy}, {Nelson},
  {Nguyen}, {Ninan}, {N{\"o}the}, {Ogaz}, {Oh}, {Parejko}, {Parley}, {Pascual},
  {Patil}, {Patil}, {Plunkett}, {Prochaska}, {Rastogi}, {Reddy Janga},
  {Sabater}, {Sakurikar}, {Seifert}, {Sherbert}, {Sherwood-Taylor}, {Shih},
  {Sick}, {Silbiger}, {Singanamalla}, {Singer}, {Sladen}, {Sooley},
  {Sornarajah}, {Streicher}, {Teuben}, {Thomas}, {Tremblay}, {Turner},
  {Terr{\'o}n}, {van Kerkwijk}, {de la Vega}, {Watkins}, {Weaver}, {Whitmore},
  {Woillez}, {Zabalza}, \& {Astropy Contributors}}]{Astropy_2018}
{Astropy Collaboration}, {Price-Whelan}, A.~M., {Sip{\H{o}}cz}, B.~M., {et~al.}
  2018, \aj, 156, 123

\bibitem[{{Bailer-Jones} {et~al.}(2021){Bailer-Jones}, {Rybizki}, {Fouesneau},
  {Demleitner}, \& {Andrae}}]{bailerjones21}
{Bailer-Jones}, C.~A.~L., {Rybizki}, J., {Fouesneau}, M., {Demleitner}, M., \&
  {Andrae}, R. 2021, \aj, 161, 147

\bibitem[{{Baumgardt}(2009)}]{Baumgardt_2009}
{Baumgardt}, H. 2009, in Globular Clusters - Guides to Galaxies, ed.
  T.~{Richtler} \& S.~{Larsen}, 387

\bibitem[{{Baumgardt} \& {Makino}(2003)}]{Baumgardt_2003}
{Baumgardt}, H. \& {Makino}, J. 2003, \mnras, 340, 227

\bibitem[{{Binney} \& {Tremaine}(2008)}]{binneytremaine08}
{Binney}, J. \& {Tremaine}, S. 2008, {Galactic Dynamics: Second Edition}

\bibitem[{{Boesgaard} {et~al.}(2015){Boesgaard}, {Lum}, \&
  {Deliyannis}}]{Boesgaard_2015}
{Boesgaard}, A.~M., {Lum}, M.~G., \& {Deliyannis}, C.~P. 2015, \apj, 799, 202

\bibitem[{Boser {et~al.}(1992)Boser, Guyon, \& Vapnik}]{Boser92}
Boser, B.~E., Guyon, I.~M., \& Vapnik, V.~N. 1992, in Proceedings of the Fifth
  Annual Workshop on Computational Learning Theory, COLT '92 (New York, NY,
  USA: Association for Computing Machinery), 144–152

\bibitem[{{Bragaglia} {et~al.}(2014){Bragaglia}, {Sneden}, {Carretta},
  {Gratton}, {Lucatello}, {Bernath}, {Brooke}, \& {Ram}}]{bragaglia14}
{Bragaglia}, A., {Sneden}, C., {Carretta}, E., {et~al.} 2014, \apj, 796, 68

\bibitem[{{Buckner} \& {Froebrich}(2014)}]{Buckner14}
{Buckner}, A. S.~M. \& {Froebrich}, D. 2014, \mnras, 444, 290

\bibitem[{{Buder} {et~al.}(2021){Buder}, {Sharma}, {Kos}, {Amarsi},
  {Nordlander}, {Lind}, {Martell}, {Asplund}, {Bland-Hawthorn}, {Casey}, {de
  Silva}, {D'Orazi}, {Freeman}, {Hayden}, {Lewis}, {Lin}, {Schlesinger},
  {Simpson}, {Stello}, {Zucker}, {Zwitter}, {Beeson}, {Buck}, {Casagrande},
  {Clark}, {{\v{C}}otar}, {da Costa}, {de Grijs}, {Feuillet}, {Horner},
  {Kafle}, {Khanna}, {Kobayashi}, {Liu}, {Montet}, {Nandakumar}, {Nataf},
  {Ness}, {Spina}, {Tepper-Garc{\'\i}a}, {Ting}, {Traven},
  {Vogrin{\v{c}}i{\v{c}}}, {Wittenmyer}, {Wyse}, {{\v{Z}}erjal}, \& {Galah
  Collaboration}}]{Buder_2021}
{Buder}, S., {Sharma}, S., {Kos}, J., {et~al.} 2021, \mnras, 506, 150

\bibitem[{{Cai} {et~al.}(2016){Cai}, {Gieles}, {Heggie}, \&
  {Varri}}]{Maxwell_2016}
{Cai}, M.~X., {Gieles}, M., {Heggie}, D.~C., \& {Varri}, A.~L. 2016, \mnras,
  455, 596

\bibitem[{{Cantat-Gaudin} {et~al.}(2020){Cantat-Gaudin}, {Anders},
  {Castro-Ginard}, {Jordi}, {Romero-G{\'o}mez}, {Soubiran}, {Casamiquela},
  {Tarricq}, {Moitinho}, {Vallenari}, {Bragaglia}, {Krone-Martins}, \&
  {Kounkel}}]{CantatGaudin20}
{Cantat-Gaudin}, T., {Anders}, F., {Castro-Ginard}, A., {et~al.} 2020, \aap,
  640, A1

\bibitem[{{Carraro} \& {Chiosi}(1994)}]{Carraro_1994}
{Carraro}, G. \& {Chiosi}, C. 1994, \aap, 287, 761

\bibitem[{{Carrera} {et~al.}(2019){Carrera}, {Pasquato}, {Vallenari},
  {Balaguer-N{\'u}{\~n}ez}, {Cantat-Gaudin}, {Mapelli}, {Bragaglia}, {Bossini},
  {Jordi}, {Galad{\'\i}-Enr{\'\i}quez}, \& {Solano}}]{Carrera_2019}
{Carrera}, R., {Pasquato}, M., {Vallenari}, A., {et~al.} 2019, \aap, 627, A119

\bibitem[{{Carretta} {et~al.}(2007){Carretta}, {Bragaglia}, \&
  {Gratton}}]{carretta07}
{Carretta}, E., {Bragaglia}, A., \& {Gratton}, R.~G. 2007, \aap, 473, 129

\bibitem[{{Chen} {et~al.}(2022){Chen}, {Ge}, {Chen}, {Bi}, {Yu}, {Yang},
  {Ferguson}, {Wu}, \& {Li}}]{Chen_2022}
{Chen}, X., {Ge}, Z., {Chen}, Y., {et~al.} 2022, \apj, 929, 124

\bibitem[{{Chen} \& {Zhao}(2020)}]{chen20}
{Chen}, Y.~Q. \& {Zhao}, G. 2020, \mnras, 495, 2673

\bibitem[{{Costa Silva} {et~al.}(2020){Costa Silva}, {Delgado Mena}, \&
  {Tsantaki}}]{Costa_Silva20}
{Costa Silva}, A.~R., {Delgado Mena}, E., \& {Tsantaki}, M. 2020, \aap, 634,
  A136

\bibitem[{{Davenport} \& {Sandquist}(2010)}]{Davenport_2010}
{Davenport}, J. R.~A. \& {Sandquist}, E.~L. 2010, \apj, 711, 559

\bibitem[{{de Grijs} \& {Parmentier}(2007)}]{deGrijs_2007}
{de Grijs}, R. \& {Parmentier}, G. 2007, \cjaa, 7, 155

\bibitem[{{de la Fuente Marcos} {et~al.}(2014){de la Fuente Marcos}, {de la
  Fuente Marcos}, \& {Reilly}}]{delaFuenteMarcos_2014}
{de la Fuente Marcos}, R., {de la Fuente Marcos}, C., \& {Reilly}, D. 2014,
  \apss, 349, 379

\bibitem[{{De Silva} {et~al.}(2015){De Silva}, {Freeman}, {Bland-Hawthorn},
  {Martell}, {de Boer}, {Asplund}, {Keller}, {Sharma}, {Zucker}, {Zwitter},
  {Anguiano}, {Bacigalupo}, {Bayliss}, {Beavis}, {Bergemann}, {Campbell},
  {Cannon}, {Carollo}, {Casagrande}, {Casey}, {Da Costa}, {D'Orazi}, {Dotter},
  {Duong}, {Heger}, {Ireland}, {Kafle}, {Kos}, {Lattanzio}, {Lewis}, {Lin},
  {Lind}, {Munari}, {Nataf}, {O'Toole}, {Parker}, {Reid}, {Schlesinger},
  {Sheinis}, {Simpson}, {Stello}, {Ting}, {Traven}, {Watson}, {Wittenmyer},
  {Yong}, \& {{\v{Z}}erjal}}]{DeSilva_2015}
{De Silva}, G.~M., {Freeman}, K.~C., {Bland-Hawthorn}, J., {et~al.} 2015,
  \mnras, 449, 2604

\bibitem[{{Delgado Mena} {et~al.}(2017){Delgado Mena}, {Tsantaki}, {Adibekyan},
  {Sousa}, {Santos}, {Gonz{\'a}lez Hern{\'a}ndez}, \&
  {Israelian}}]{DelgadoMena17}
{Delgado Mena}, E., {Tsantaki}, M., {Adibekyan}, V.~Z., {et~al.} 2017, \aap,
  606, A94

\bibitem[{{Friel}(1995)}]{friel95}
{Friel}, E.~D. 1995, \araa, 33, 381

\bibitem[{{Fujii} \& {Baba}(2012)}]{fujii12}
{Fujii}, M.~S. \& {Baba}, J. 2012, \mnras, 427, L16

\bibitem[{{Fukushige} \& {Heggie}(2000)}]{Fukushige_2000}
{Fukushige}, T. \& {Heggie}, D.~C. 2000, \mnras, 318, 753

\bibitem[{{Gaia Collaboration} {et~al.}(2021){Gaia Collaboration}, {Brown},
  {Vallenari}, {Prusti}, {de Bruijne}, {Babusiaux}, {Biermann}, {Creevey},
  {Evans}, {Eyer}, {Hutton}, {Jansen}, {Jordi}, {Klioner}, {Lammers},
  {Lindegren}, {Luri}, {Mignard}, {Panem}, {Pourbaix}, {Randich}, {Sartoretti},
  {Soubiran}, {Walton}, {Arenou}, {Bailer-Jones}, {Bastian}, {Cropper},
  {Drimmel}, {Katz}, {Lattanzi}, {van Leeuwen}, {Bakker}, {Cacciari},
  {Casta{\~n}eda}, {De Angeli}, {Ducourant}, {Fabricius}, {Fouesneau},
  {Fr{\'e}mat}, {Guerra}, {Guerrier}, {Guiraud}, {Jean-Antoine Piccolo},
  {Masana}, {Messineo}, {Mowlavi}, {Nicolas}, {Nienartowicz}, {Pailler},
  {Panuzzo}, {Riclet}, {Roux}, {Seabroke}, {Sordo}, {Tanga}, {Th{\'e}venin},
  {Gracia-Abril}, {Portell}, {Teyssier}, {Altmann}, {Andrae}, {Bellas-Velidis},
  {Benson}, {Berthier}, {Blomme}, {Brugaletta}, {Burgess}, {Busso}, {Carry},
  {Cellino}, {Cheek}, {Clementini}, {Damerdji}, {Davidson}, {Delchambre},
  {Dell'Oro}, {Fern{\'a}ndez-Hern{\'a}ndez}, {Galluccio}, {Garc{\'\i}a-Lario},
  {Garcia-Reinaldos}, {Gonz{\'a}lez-N{\'u}{\~n}ez}, {Gosset}, {Haigron},
  {Halbwachs}, {Hambly}, {Harrison}, {Hatzidimitriou}, {Heiter},
  {Hern{\'a}ndez}, {Hestroffer}, {Hodgkin}, {Holl}, {Jan{\ss}en}, {Jevardat de
  Fombelle}, {Jordan}, {Krone-Martins}, {Lanzafame}, {L{\"o}ffler}, {Lorca},
  {Manteiga}, {Marchal}, {Marrese}, {Moitinho}, {Mora}, {Muinonen}, {Osborne},
  {Pancino}, {Pauwels}, {Petit}, {Recio-Blanco}, {Richards}, {Riello},
  {Rimoldini}, {Robin}, {Roegiers}, {Rybizki}, {Sarro}, {Siopis}, {Smith},
  {Sozzetti}, {Ulla}, {Utrilla}, {van Leeuwen}, {van Reeven}, {Abbas}, {Abreu
  Aramburu}, {Accart}, {Aerts}, {Aguado}, {Ajaj}, {Altavilla}, {{\'A}lvarez},
  {{\'A}lvarez Cid-Fuentes}, {Alves}, {Anderson}, {Anglada Varela}, {Antoja},
  {Audard}, {Baines}, {Baker}, {Balaguer-N{\'u}{\~n}ez}, {Balbinot}, {Balog},
  {Barache}, {Barbato}, {Barros}, {Barstow}, {Bartolom{\'e}}, {Bassilana},
  {Bauchet}, {Baudesson-Stella}, {Becciani}, {Bellazzini}, {Bernet}, {Bertone},
  {Bianchi}, {Blanco-Cuaresma}, {Boch}, {Bombrun}, {Bossini}, {Bouquillon},
  {Bragaglia}, {Bramante}, {Breedt}, {Bressan}, {Brouillet}, {Bucciarelli},
  {Burlacu}, {Busonero}, {Butkevich}, {Buzzi}, {Caffau}, {Cancelliere},
  {C{\'a}novas}, {Cantat-Gaudin}, {Carballo}, {Carlucci}, {Carnerero},
  {Carrasco}, {Casamiquela}, {Castellani}, {Castro-Ginard}, {Castro Sampol},
  {Chaoul}, {Charlot}, {Chemin}, {Chiavassa}, {Cioni}, {Comoretto}, {Cooper},
  {Cornez}, {Cowell}, {Crifo}, {Crosta}, {Crowley}, {Dafonte}, {Dapergolas},
  {David}, {David}, {de Laverny}, {De Luise}, {De March}, {De Ridder}, {de
  Souza}, {de Teodoro}, {de Torres}, {del Peloso}, {del Pozo}, {Delbo},
  {Delgado}, {Delgado}, {Delisle}, {Di Matteo}, {Diakite}, {Diener},
  {Distefano}, {Dolding}, {Eappachen}, {Edvardsson}, {Enke}, {Esquej}, {Fabre},
  {Fabrizio}, {Faigler}, {Fedorets}, {Fernique}, {Fienga}, {Figueras},
  {Fouron}, {Fragkoudi}, {Fraile}, {Franke}, {Gai}, {Garabato},
  {Garcia-Gutierrez}, {Garc{\'\i}a-Torres}, {Garofalo}, {Gavras}, {Gerlach},
  {Geyer}, {Giacobbe}, {Gilmore}, {Girona}, {Giuffrida}, {Gomel}, {Gomez},
  {Gonzalez-Santamaria}, {Gonz{\'a}lez-Vidal}, {Granvik},
  {Guti{\'e}rrez-S{\'a}nchez}, {Guy}, {Hauser}, {Haywood}, {Helmi}, {Hidalgo},
  {Hilger}, {H{\l}adczuk}, {Hobbs}, {Holland}, {Huckle}, {Jasniewicz},
  {Jonker}, {Juaristi Campillo}, {Julbe}, {Karbevska}, {Kervella}, {Khanna},
  {Kochoska}, {Kontizas}, {Kordopatis}, {Korn}, {Kostrzewa-Rutkowska},
  {Kruszy{\'n}ska}, {Lambert}, {Lanza}, {Lasne}, {Le Campion}, {Le Fustec},
  {Lebreton}, {Lebzelter}, {Leccia}, {Leclerc}, {Lecoeur-Taibi}, {Liao},
  {Licata}, {Lindstr{\o}m}, {Lister}, {Livanou}, {Lobel}, {Madrero Pardo},
  {Managau}, {Mann}, {Marchant}, {Marconi}, {Marcos Santos}, {Marinoni},
  {Marocco}, {Marshall}, {Martin Polo}, {Mart{\'\i}n-Fleitas}, {Masip},
  {Massari}, {Mastrobuono-Battisti}, {Mazeh}, {McMillan}, {Messina},
  {Michalik}, {Millar}, {Mints}, {Molina}, {Molinaro}, {Moln{\'a}r},
  {Montegriffo}, {Mor}, {Morbidelli}, {Morel}, {Morris}, {Mulone}, {Munoz},
  {Muraveva}, {Murphy}, {Musella}, {Noval}, {Ord{\'e}novic}, {Orr{\`u}},
  {Osinde}, {Pagani}, {Pagano}, {Palaversa}, {Palicio}, {Panahi}, {Pawlak},
  {Pe{\~n}alosa Esteller}, {Penttil{\"a}}, {Piersimoni}, {Pineau}, {Plachy},
  {Plum}, {Poggio}, {Poretti}, {Poujoulet}, {Pr{\v{s}}a}, {Pulone}, {Racero},
  {Ragaini}, {Rainer}, {Raiteri}, {Rambaux}, {Ramos}, {Ramos-Lerate}, {Re
  Fiorentin}, {Regibo}, {Reyl{\'e}}, {Ripepi}, {Riva}, {Rixon}, {Robichon},
  {Robin}, {Roelens}, {Rohrbasser}, {Romero-G{\'o}mez}, {Rowell}, {Royer},
  {Rybicki}, {Sadowski}, {Sagrist{\`a} Sell{\'e}s}, {Sahlmann}, {Salgado},
  {Salguero}, {Samaras}, {Sanchez Gimenez}, {Sanna}, {Santove{\~n}a},
  {Sarasso}, {Schultheis}, {Sciacca}, {Segol}, {Segovia}, {S{\'e}gransan},
  {Semeux}, {Shahaf}, {Siddiqui}, {Siebert}, {Siltala}, {Slezak}, {Smart},
  {Solano}, {Solitro}, {Souami}, {Souchay}, {Spagna}, {Spoto}, {Steele},
  {Steidelm{\"u}ller}, {Stephenson}, {S{\"u}veges}, {Szabados}, {Szegedi-Elek},
  {Taris}, {Tauran}, {Taylor}, {Teixeira}, {Thuillot}, {Tonello}, {Torra},
  {Torra}, {Turon}, {Unger}, {Vaillant}, {van Dillen}, {Vanel}, {Vecchiato},
  {Viala}, {Vicente}, {Voutsinas}, {Weiler}, {Wevers}, {Wyrzykowski}, {Yoldas},
  {Yvard}, {Zhao}, {Zorec}, {Zucker}, {Zurbach}, \& {Zwitter}}]{gaiadr3}
{Gaia Collaboration}, {Brown}, A.~G.~A., {Vallenari}, A., {et~al.} 2021, \aap,
  649, A1

\bibitem[{{Gaia Collaboration} {et~al.}(2023){Gaia Collaboration},
  {Recio-Blanco}, {Kordopatis}, {de Laverny}, {Palicio}, {Spagna}, {Spina},
  {Katz}, {Re Fiorentin}, {Poggio}, {McMillan}, {Vallenari}, {Lattanzi},
  {Seabroke}, {Casamiquela}, {Bragaglia}, {Antoja}, {Bailer-Jones},
  {Schultheis}, {Andrae}, {Fouesneau}, {Cropper}, {Cantat-Gaudin}, {Bijaoui},
  {Heiter}, {Brown}, {Prusti}, {de Bruijne}, {Arenou}, {Babusiaux}, {Biermann},
  {Creevey}, {Ducourant}, {Evans}, {Eyer}, {Guerra}, {Hutton}, {Jordi},
  {Klioner}, {Lammers}, {Lindegren}, {Luri}, {Mignard}, {Panem}, {Pourbaix},
  {Randich}, {Sartoretti}, {Soubiran}, {Tanga}, {Walton}, {Bastian}, {Drimmel},
  {Jansen}, {van Leeuwen}, {Bakker}, {Cacciari}, {Casta{\~n}eda}, {De Angeli},
  {Fabricius}, {Fr{\'e}mat}, {Galluccio}, {Guerrier}, {Masana}, {Messineo},
  {Mowlavi}, {Nicolas}, {Nienartowicz}, {Pailler}, {Panuzzo}, {Riclet}, {Roux},
  {Sordo}, {Th{\'e}venin}, {Gracia-Abril}, {Portell}, {Teyssier}, {Altmann},
  {Audard}, {Bellas-Velidis}, {Benson}, {Berthier}, {Blomme}, {Burgess},
  {Busonero}, {Busso}, {C{\'a}novas}, {Carry}, {Cellino}, {Cheek},
  {Clementini}, {Damerdji}, {Davidson}, {de Teodoro}, {Nu{\~n}ez Campos},
  {Delchambre}, {Dell'Oro}, {Esquej}, {Fern{\'a}ndez-Hern{\'a}ndez}, {Fraile},
  {Garabato}, {Garc{\'\i}a-Lario}, {Gosset}, {Haigron}, {Halbwachs}, {Hambly},
  {Harrison}, {Hern{\'a}ndez}, {Hestroffer}, {Hodgkin}, {Holl}, {Jan{\ss}en},
  {Jevardat de Fombelle}, {Jordan}, {Krone-Martins}, {Lanzafame},
  {L{\"o}ffler}, {Marchal}, {Marrese}, {Moitinho}, {Muinonen}, {Osborne},
  {Pancino}, {Pauwels}, {Reyl{\'e}}, {Riello}, {Rimoldini}, {Roegiers},
  {Rybizki}, {Sarro}, {Siopis}, {Smith}, {Sozzetti}, {Utrilla}, {van Leeuwen},
  {Abbas}, {{\'A}brah{\'a}m}, {Abreu Aramburu}, {Aerts}, {Aguado}, {Ajaj},
  {Aldea-Montero}, {Altavilla}, {{\'A}lvarez}, {Alves}, {Anders}, {Anderson},
  {Anglada Varela}, {Baines}, {Baker}, {Balaguer-N{\'u}{\~n}ez}, {Balbinot},
  {Balog}, {Barache}, {Barbato}, {Barros}, {Barstow}, {Bartolom{\'e}},
  {Bassilana}, {Bauchet}, {Becciani}, {Bellazzini}, {Berihuete}, {Bernet},
  {Bertone}, {Bianchi}, {Binnenfeld}, {Blanco-Cuaresma}, {Boch}, {Bombrun},
  {Bossini}, {Bouquillon}, {Bramante}, {Breedt}, {Bressan}, {Brouillet},
  {Brugaletta}, {Bucciarelli}, {Burlacu}, {Butkevich}, {Buzzi}, {Caffau},
  {Cancelliere}, {Carballo}, {Carlucci}, {Carnerero}, {Carrasco}, {Castellani},
  {Castro-Ginard}, {Chaoul}, {Charlot}, {Chemin}, {Chiaramida}, {Chiavassa},
  {Chornay}, {Comoretto}, {Contursi}, {Cooper}, {Cornez}, {Cowell}, {Crifo},
  {Crosta}, {Crowley}, {Dafonte}, {Dapergolas}, {David}, {De Luise}, {De
  March}, {De Ridder}, {de Souza}, {de Torres}, {del Peloso}, {del Pozo},
  {Delbo}, {Delgado}, {Delisle}, {Demouchy}, {Dharmawardena}, {Di Matteo},
  {Diakite}, {Diener}, {Distefano}, {Dolding}, {Edvardsson}, {Enke}, {Fabre},
  {Fabrizio}, {Faigler}, {Fedorets}, {Fernique}, {Figueras}, {Fournier},
  {Fouron}, {Fragkoudi}, {Gai}, {Garcia-Gutierrez}, {Garcia-Reinaldos},
  {Garc{\'\i}a-Torres}, {Garofalo}, {Gavel}, {Gavras}, {Gerlach}, {Geyer},
  {Giacobbe}, {Gilmore}, {Girona}, {Giuffrida}, {Gomel}, {Gomez},
  {Gonz{\'a}lez-N{\'u}{\~n}ez}, {Gonz{\'a}lez-Santamar{\'\i}a},
  {Gonz{\'a}lez-Vidal}, {Granvik}, {Guillout}, {Guiraud},
  {Guti{\'e}rrez-S{\'a}nchez}, {Guy}, {Hatzidimitriou}, {Hauser}, {Haywood},
  {Helmer}, {Helmi}, {Sarmiento}, {Hidalgo}, {H{\l}adczuk}, {Hobbs}, {Holland},
  {Huckle}, {Jardine}, {Jasniewicz}, {Jean-Antoine Piccolo},
  {Jim{\'e}nez-Arranz}, {Juaristi Campillo}, {Julbe}, {Karbevska}, {Kervella},
  {Khanna}, {Korn}, {K{\'o}sp{\'a}l}, {Kostrzewa-Rutkowska}, {Kruszy{\'n}ska},
  {Kun}, {Laizeau}, {Lambert}, {Lanza}, {Lasne}, {Le Campion}, {Lebreton},
  {Lebzelter}, {Leccia}, {Leclerc}, {Lecoeur-Taibi}, {Liao}, {Licata},
  {Lindstr{\o}m}, {Lister}, {Livanou}, {Lobel}, {Lorca}, {Loup}, {Madrero
  Pardo}, {Magdaleno Romeo}, {Managau}, {Mann}, {Manteiga}, {Marchant},
  {Marconi}, {Marcos}, {Marcos Santos}, {Mar{\'\i}n Pina}, {Marinoni},
  {Marocco}, {Marshall}, {Martin Polo}, {Mart{\'\i}n-Fleitas}, {Marton},
  {Mary}, {Masip}, {Massari}, {Mastrobuono-Battisti}, {Mazeh}, {Messina},
  {Michalik}, {Millar}, {Mints}, {Molina}, {Molinaro}, {Moln{\'a}r}, {Monari},
  {Mongui{\'o}}, {Montegriffo}, {Montero}, {Mor}, {Mora}, {Morbidelli},
  {Morel}, {Morris}, {Muraveva}, {Murphy}, {Musella}, {Nagy}, {Noval},
  {Oca{\~n}a}, {Ogden}, {Ordenovic}, {Osinde}, {Pagani}, {Pagano}, {Palaversa},
  {Pallas-Quintela}, {Panahi}, {Payne-Wardenaar}, {Pe{\~n}alosa Esteller},
  {Penttil{\"a}}, {Pichon}, {Piersimoni}, {Pineau}, {Plachy}, {Plum},
  {Pr{\v{s}}a}, {Pulone}, {Racero}, {Ragaini}, {Rainer}, {Raiteri}, {Ramos},
  {Ramos-Lerate}, {Regibo}, {Richards}, {Rios Diaz}, {Ripepi}, {Riva}, {Rix},
  {Rixon}, {Robichon}, {Robin}, {Robin}, {Roelens}, {Rogues}, {Rohrbasser},
  {Romero-G{\'o}mez}, {Rowell}, {Royer}, {Ruz Mieres}, {Rybicki}, {Sadowski},
  {S{\'a}ez N{\'u}{\~n}ez}, {Sagrist{\`a} Sell{\'e}s}, {Sahlmann}, {Salguero},
  {Samaras}, {Sanchez Gimenez}, {Sanna}, {Santove{\~n}a}, {Sarasso}, {Sciacca},
  {Segol}, {Segovia}, {S{\'e}gransan}, {Semeux}, {Shahaf}, {Siddiqui},
  {Siebert}, {Siltala}, {Silvelo}, {Slezak}, {Slezak}, {Smart}, {Snaith},
  {Solano}, {Solitro}, {Souami}, {Souchay}, {Spoto}, {Steele},
  {Steidelm{\"u}ller}, {Stephenson}, {S{\"u}veges}, {Surdej}, {Szabados},
  {Szegedi-Elek}, {Taris}, {Taylor}, {Teixeira}, {Tolomei}, {Tonello}, {Torra},
  {Torra}, {Torralba Elipe}, {Trabucchi}, {Tsounis}, {Turon}, {Ulla}, {Unger},
  {Vaillant}, {van Dillen}, {van Reeven}, {Vanel}, {Vecchiato}, {Viala},
  {Vicente}, {Voutsinas}, {Weiler}, {Wevers}, {Wyrzykowski}, {Yoldas}, {Yvard},
  {Zhao}, {Zorec}, {Zucker}, \& {Zwitter}}]{Gaia23_chemical_cartography}
{Gaia Collaboration}, {Recio-Blanco}, A., {Kordopatis}, G., {et~al.} 2023,
  \aap, 674, A38

\bibitem[{{Gao} {et~al.}(2014){Gao}, {Chen}, \& {Hou}}]{Gao_2014}
{Gao}, X.-h., {Chen}, L., \& {Hou}, Z.-j. 2014, \caa, 38, 257

\bibitem[{{Geller} {et~al.}(2015){Geller}, {Latham}, \&
  {Mathieu}}]{Geller_2015}
{Geller}, A.~M., {Latham}, D.~W., \& {Mathieu}, R.~D. 2015, \aj, 150, 97

\bibitem[{{Gieles} {et~al.}(2007){Gieles}, {Athanassoula}, \& {Portegies
  Zwart}}]{Gieles_2007}
{Gieles}, M., {Athanassoula}, E., \& {Portegies Zwart}, S.~F. 2007, \mnras,
  376, 809

\bibitem[{{Gieles} {et~al.}(2006){Gieles}, {Portegies Zwart}, {Baumgardt},
  {Athanassoula}, {Lamers}, {Sipior}, \& {Leenaarts}}]{Gieles_2006}
{Gieles}, M., {Portegies Zwart}, S.~F., {Baumgardt}, H., {et~al.} 2006, \mnras,
  371, 793

\bibitem[{{Gieles} \& {Renaud}(2016)}]{gieles16}
{Gieles}, M. \& {Renaud}, F. 2016, \mnras, 463, L103

\bibitem[{{GRAVITY Collaboration} {et~al.}(2020){GRAVITY Collaboration},
  {Abuter}, {Amorim}, {Baub{\"o}ck}, {Berger}, {Bonnet}, {Brandner}, {Cardoso},
  {Cl{\'e}net}, {de Zeeuw}, {Dexter}, {Eckart}, {Eisenhauer}, {F{\"o}rster
  Schreiber}, {Garcia}, {Gao}, {Gendron}, {Genzel}, {Gillessen}, {Habibi},
  {Haubois}, {Henning}, {Hippler}, {Horrobin}, {Jim{\'e}nez-Rosales}, {Jochum},
  {Jocou}, {Kaufer}, {Kervella}, {Lacour}, {Lapeyr{\`e}re}, {Le Bouquin},
  {L{\'e}na}, {Nowak}, {Ott}, {Paumard}, {Perraut}, {Perrin}, {Pfuhl},
  {Rodr{\'\i}guez-Coira}, {Shangguan}, {Scheithauer}, {Stadler}, {Straub},
  {Straubmeier}, {Sturm}, {Tacconi}, {Vincent}, {von Fellenberg}, {Waisberg},
  {Widmann}, {Wieprecht}, {Wiezorrek}, {Woillez}, {Yazici}, \&
  {Zins}}]{gravity20}
{GRAVITY Collaboration}, {Abuter}, R., {Amorim}, A., {et~al.} 2020, \aap, 636,
  L5

\bibitem[{{Grebel}(2000)}]{Grebel_2000}
{Grebel}, E.~K. 2000, in Astronomical Society of the Pacific Conference Series,
  Vol. 211, Massive Stellar Clusters, ed. A.~{Lan{\c{c}}on} \& C.~M. {Boily},
  262

\bibitem[{{Gustafsson} {et~al.}(2016){Gustafsson}, {Church}, {Davies}, \&
  {Rickman}}]{gustafsson16}
{Gustafsson}, B., {Church}, R.~P., {Davies}, M.~B., \& {Rickman}, H. 2016,
  \aap, 593, A85

\bibitem[{{Howes} {et~al.}(2019){Howes}, {Lindegren}, {Feltzing}, {Church}, \&
  {Bensby}}]{Howes_2019}
{Howes}, L.~M., {Lindegren}, L., {Feltzing}, S., {Church}, R.~P., \& {Bensby},
  T. 2019, \aap, 622, A27

\bibitem[{Hunter(2007)}]{Hunter2007}
Hunter, J.~D. 2007, Computing in Science \& Engineering, 9, 90

\bibitem[{{J{\'\i}lkov{\'a}} {et~al.}(2012){J{\'\i}lkov{\'a}}, {Carraro},
  {Jungwiert}, \& {Minchev}}]{Jílkova_2012}
{J{\'\i}lkov{\'a}}, L., {Carraro}, G., {Jungwiert}, B., \& {Minchev}, I. 2012,
  \aap, 541, A64

\bibitem[{{J{\"o}nsson} {et~al.}(2020){J{\"o}nsson}, {Holtzman}, {Allende
  Prieto}, {Cunha}, {Garc{\'\i}a-Hern{\'a}ndez}, {Hasselquist}, {Masseron},
  {Osorio}, {Shetrone}, {Smith}, {Stringfellow}, {Bizyaev}, {Edvardsson},
  {Majewski}, {M{\'e}sz{\'a}ros}, {Souto}, {Zamora}, {Beaton}, {Bovy}, {Donor},
  {Pinsonneault}, {Poovelil}, \& {Sobeck}}]{Jonsson_2020}
{J{\"o}nsson}, H., {Holtzman}, J.~A., {Allende Prieto}, C., {et~al.} 2020, \aj,
  160, 120

\bibitem[{{Khoperskov} {et~al.}(2018){Khoperskov}, {Mastrobuono-Battisti}, {Di
  Matteo}, \& {Haywood}}]{Khoperskov_2018}
{Khoperskov}, S., {Mastrobuono-Battisti}, A., {Di Matteo}, P., \& {Haywood}, M.
  2018, \aap, 620, A154

\bibitem[{{Kolmogorov}(1933)}]{kolmogorov33}
{Kolmogorov}, A. 1933, Giornale dell' Istituto Italiano degli Attuari, 4, 83

\bibitem[{{Kordopatis} {et~al.}(2023){Kordopatis}, {Schultheis}, {McMillan},
  {Palicio}, {de Laverny}, {Recio-Blanco}, {Creevey}, {{\'A}lvarez}, {Andrae},
  {Poggio}, {Spitoni}, {Contursi}, {Zhao}, {Oreshina-Slezak}, {Ordenovic}, \&
  {Bijaoui}}]{Kordopatis_2023}
{Kordopatis}, G., {Schultheis}, M., {McMillan}, P.~J., {et~al.} 2023, \aap,
  669, A104

\bibitem[{{Kruijssen}(2012)}]{Kruijssen_2012}
{Kruijssen}, J.~M.~D. 2012, \mnras, 426, 3008

\bibitem[{{Kubryk} {et~al.}(2013){Kubryk}, {Prantzos}, \&
  {Athanassoula}}]{Kubryk_2013}
{Kubryk}, M., {Prantzos}, N., \& {Athanassoula}, E. 2013, arXiv e-prints,
  arXiv:1309.2437

\bibitem[{{Lamers} {et~al.}(2005){Lamers}, {Gieles}, \& {Portegies
  Zwart}}]{Lamers_2005}
{Lamers}, H.~J.~G.~L.~M., {Gieles}, M., \& {Portegies Zwart}, S.~F. 2005, \aap,
  429, 173

\bibitem[{{Li} {et~al.}(2017){Li}, {Gnedin}, {Gnedin}, {Meng}, {Semenov}, \&
  {Kravtsov}}]{li17}
{Li}, H., {Gnedin}, O.~Y., {Gnedin}, N.~Y., {et~al.} 2017, \apj, 834, 69

\bibitem[{{Loebman} {et~al.}(2016){Loebman}, {Debattista}, {Nidever}, {Hayden},
  {Holtzman}, {Clarke}, {Ro{\v{s}}kar}, \& {Valluri}}]{Loebman_2016}
{Loebman}, S.~R., {Debattista}, V.~P., {Nidever}, D.~L., {et~al.} 2016, \apjl,
  818, L6

\bibitem[{{Magrini} {et~al.}(2017){Magrini}, {Randich}, {Kordopatis},
  {Prantzos}, {Romano}, {Chieffi}, {Limongi}, {Fran{\c{c}}ois}, {Pancino},
  {Friel}, {Bragaglia}, {Tautvai{\v{s}}ien{\.{e}}}, {Spina}, {Overbeek},
  {Cantat-Gaudin}, {Donati}, {Vallenari}, {Sordo}, {Jim{\'e}nez-Esteban},
  {Tang}, {Drazdauskas}, {Sousa}, {Duffau}, {Jofr{\'e}}, {Gilmore}, {Feltzing},
  {Alfaro}, {Bensby}, {Flaccomio}, {Koposov}, {Lanzafame}, {Smiljanic}, {Bayo},
  {Carraro}, {Casey}, {Costado}, {Damiani}, {Franciosini}, {Hourihane},
  {Lardo}, {Lewis}, {Monaco}, {Morbidelli}, {Sacco}, {Sbordone}, {Worley}, \&
  {Zaggia}}]{magrini17}
{Magrini}, L., {Randich}, S., {Kordopatis}, G., {et~al.} 2017, \aap, 603, A2

\bibitem[{{Magrini} {et~al.}(2023){Magrini}, {Viscasillas V{\'a}zquez},
  {Spina}, {Randich}, {Romano}, {Franciosini}, {Recio-Blanco}, {Nordlander},
  {D'Orazi}, {Baratella}, {Smiljanic}, {Dantas}, {Pasquini}, {Spitoni},
  {Casali}, {Van der Swaelmen}, {Bensby}, {Stonkute}, {Feltzing}, {Sacco},
  {Bragaglia}, {Pancino}, {Heiter}, {Biazzo}, {Gilmore}, {Bergemann},
  {Tautvai{\v{s}}ien{\.{e}}}, {Worley}, {Hourihane}, {Gonneau}, \&
  {Morbidelli}}]{magrini23}
{Magrini}, L., {Viscasillas V{\'a}zquez}, C., {Spina}, L., {et~al.} 2023, \aap,
  669, A119

\bibitem[{{McMillan}(2017)}]{mcmillan17}
{McMillan}, P.~J. 2017, \mnras, 465, 76

\bibitem[{{Meng} \& {Gnedin}(2022)}]{Meng_2022}
{Meng}, X. \& {Gnedin}, O.~Y. 2022, \mnras, 515, 1065

\bibitem[{{Mieske} \& {Baumgardt}(2007)}]{Mieske_2007}
{Mieske}, S. \& {Baumgardt}, H. 2007, \aap, 475, 851

\bibitem[{{Moitinho}(2010)}]{moitinho10}
{Moitinho}, A. 2010, in Star Clusters: Basic Galactic Building Blocks
  Throughout Time and Space, ed. R.~{de Grijs} \& J.~R.~D. {L{\'e}pine}, Vol.
  266, 106--116

\bibitem[{{Moyano Loyola} \& {Hurley}(2013)}]{Moyano_Loyola_2013}
{Moyano Loyola}, G. R.~I. \& {Hurley}, J.~R. 2013, \mnras, 434, 2509

\bibitem[{{Palla} {et~al.}(2022){Palla}, {Santos-Peral}, {Recio-Blanco}, \&
  {Matteucci}}]{palla22}
{Palla}, M., {Santos-Peral}, P., {Recio-Blanco}, A., \& {Matteucci}, F. 2022,
  \aap, 663, A125

\bibitem[{pandas~development team(2020)}]{pandas2020}
pandas~development team, T. 2020, pandas-dev/pandas: Pandas

\bibitem[{Pedregosa {et~al.}(2011)Pedregosa, Varoquaux, Gramfort, Michel,
  Thirion, Grisel, Blondel, Prettenhofer, Weiss, Dubourg, Vanderplas, Passos,
  Cournapeau, Brucher, Perrot, \& Duchesnay}]{scikit-learn11}
Pedregosa, F., Varoquaux, G., Gramfort, A., {et~al.} 2011, Journal of Machine
  Learning Research, 12, 2825

\bibitem[{{Portegies Zwart} {et~al.}(2002){Portegies Zwart}, {Makino},
  {McMillan}, \& {Hut}}]{Portegies_2002}
{Portegies Zwart}, S.~F., {Makino}, J., {McMillan}, S. L.~W., \& {Hut}, P.
  2002, \apj, 565, 265

\bibitem[{{Recio-Blanco} {et~al.}(2016){Recio-Blanco}, {de Laverny}, {Allende
  Prieto}, {Fustes}, {Manteiga}, {Arcay}, {Bijaoui}, {Dafonte}, {Ordenovic}, \&
  {Ordo{\~n}ez Blanco}}]{Recio-Blanco_2016}
{Recio-Blanco}, A., {de Laverny}, P., {Allende Prieto}, C., {et~al.} 2016,
  \aap, 585, A93

\bibitem[{{Recio-Blanco} {et~al.}(2022){Recio-Blanco}, {de Laverny}, {Palicio},
  {Kordopatis}, {{\'A}lvarez}, {Schultheis}, {Contursi}, {Zhao}, {Torralba
  Elipe}, {Ordenovic}, {Manteiga}, {Dafonte}, {Oreshina-Slezak}, {Bijaoui},
  {Fremat}, {Seabroke}, {Pailler}, {Spitoni}, {Poggio}, {Creevey}, {Abreu
  Aramburu}, {Accart}, {Andrae}, {Bailer-Jones}, {Bellas-Velidis}, {Brouillet},
  {Brugaletta}, {Burlacu}, {Carballo}, {Casamiquela}, {Chiavassa}, {Cooper},
  {Dapergolas}, {Delchambre}, {Dharmawardena}, {Drimmel}, {Edvardsson},
  {Fouesneau}, {Garabato}, {Garcia-Lario}, {Garcia-Torres}, {Gavel}, {Gomez},
  {Gonzalez-Santamaria}, {Hatzidimitriou}, {Heiter}, {Jean-Antoine Piccolo},
  {Kontizas}, {Korn}, {Lanzafame}, {Lebreton}, {Le Fustec}, {Licata},
  {Lindstrom}, {Livanou}, {Lobel}, {Lorca}, {Magdaleno Romeo}, {Marocco},
  {Marshall}, {Mary}, {Nicolas}, {Pallas-Quintela}, {Panem}, {Pichon},
  {Riclet}, {Robin}, {Rybizki}, {Santovena}, {Silvelo}, {Smart}, {Sarro},
  {Sordo}, {Soubiran}, {Suvege}, {Ulla}, {Vallenari}, {Zorec}, {Utrilla}, \&
  {Bakker}}]{Recio-Blanco_2022}
{Recio-Blanco}, A., {de Laverny}, P., {Palicio}, P.~A., {et~al.} 2022, arXiv
  e-prints, arXiv:2206.05541

\bibitem[{{Renaud}(2018)}]{Renaud18}
{Renaud}, F. 2018, \nar, 81, 1

\bibitem[{{Russeil} {et~al.}(2017){Russeil}, {Zavagno}, {M{\`e}ge}, {Poulin},
  {Molinari}, \& {Cambresy}}]{russel17}
{Russeil}, D., {Zavagno}, A., {M{\`e}ge}, P., {et~al.} 2017, \aap, 601, L5

\bibitem[{Seabold \& Perktold(2010)}]{seabold2010statsmodels}
Seabold, S. \& Perktold, J. 2010, in 9th Python in Science Conference

\bibitem[{{Sellwood} \& {Binney}(2002)}]{Sellwood_2002}
{Sellwood}, J.~A. \& {Binney}, J.~J. 2002, \mnras, 336, 785

\bibitem[{Smirnov(1939)}]{smirnov1939estimate}
Smirnov, N.~V. 1939, Bulletin Moscow University, 2, 3

\bibitem[{{Spina} {et~al.}(2021){Spina}, {Ting}, {De Silva}, {Frankel},
  {Sharma}, {Cantat-Gaudin}, {Joyce}, {Stello}, {Karakas}, {Asplund},
  {Nordlander}, {Casagrande}, {D'Orazi}, {Casey}, {Cottrell},
  {Tepper-Garc{\'\i}a}, {Baratella}, {Kos}, {{\v{C}}otar}, {Bland-Hawthorn},
  {Buder}, {Freeman}, {Hayden}, {Lewis}, {Lin}, {Lind}, {Martell},
  {Schlesinger}, {Simpson}, {Zucker}, \& {Zwitter}}]{Spina_2021}
{Spina}, L., {Ting}, Y.~S., {De Silva}, G.~M., {et~al.} 2021, \mnras, 503, 3279

\bibitem[{{Taylor}(2005)}]{Taylor2005}
{Taylor}, M.~B. 2005, in Astronomical Society of the Pacific Conference Series,
  Vol. 347, Astronomical Data Analysis Software and Systems XIV, ed.
  P.~{Shopbell}, M.~{Britton}, \& R.~{Ebert}, 29

\bibitem[{{Terlevich}(1987)}]{Terlevich_1987}
{Terlevich}, E. 1987, \mnras, 224, 193

\bibitem[{{Trick} {et~al.}(2019){Trick}, {Coronado}, \& {Rix}}]{trick19}
{Trick}, W.~H., {Coronado}, J., \& {Rix}, H.-W. 2019, \mnras, 484, 3291

\bibitem[{{van den Bergh} \& {McClure}(1980)}]{van_den_Bergh_1980}
{van den Bergh}, S. \& {McClure}, R.~D. 1980, \aap, 88, 360

\bibitem[{Virtanen {et~al.}(2020)Virtanen, Gommers, Oliphant, Haberland, Reddy,
  Cournapeau, Burovski, Peterson, Weckesser, Bright, {van der Walt}, Brett,
  Wilson, Millman, Mayorov, Nelson, Jones, Kern, Larson, Carey, Polat, Feng,
  Moore, {VanderPlas}, Laxalde, Perktold, Cimrman, Henriksen, Quintero, Harris,
  Archibald, Ribeiro, Pedregosa, {van Mulbregt}, \& {SciPy 1.0
  Contributors}}]{SciPy20}
Virtanen, P., Gommers, R., Oliphant, T.~E., {et~al.} 2020, Nature Methods, 17,
  261

\bibitem[{{Viscasillas V{\'a}zquez} {et~al.}(2022){Viscasillas V{\'a}zquez},
  {Magrini}, {Casali}, {Tautvai{\v{s}}ien{\.{e}}}, {Spina}, {Van der Swaelmen},
  {Randich}, {Bensby}, {Bragaglia}, {Friel}, {Feltzing}, {Sacco}, {Turchi},
  {Jim{\'e}nez-Esteban}, {D'Orazi}, {Delgado-Mena}, {Mikolaitis},
  {Drazdauskas}, {Minkevi{\v{c}}i{\={u}}t{\.{e}}}, {Stonkut{\.{e}}},
  {Bagdonas}, {Montes}, {Guiglion}, {Baratella}, {Tabernero}, {Gilmore},
  {Alfaro}, {Francois}, {Korn}, {Smiljanic}, {Bergemann}, {Franciosini},
  {Gonneau}, {Hourihane}, {Worley}, \& {Zaggia}}]{Viscasillas_2022}
{Viscasillas V{\'a}zquez}, C., {Magrini}, L., {Casali}, G., {et~al.} 2022,
  \aap, 660, A135

\bibitem[{Waskom(2021)}]{Waskom2021}
Waskom, M.~L. 2021, Journal of Open Source Software, 6, 3021

\bibitem[{Wu {et~al.}(2004)Wu, Lin, \& Weng}]{Wu04}
Wu, T.-F., Lin, C.-J., \& Weng, R.~C. 2004, J. Mach. Learn. Res., 5, 975–1005

\bibitem[{{Yoon} {et~al.}(2019){Yoon}, {Im}, {Lee}, {Lee}, \&
  {Lim}}]{Yoon_2019}
{Yoon}, Y., {Im}, M., {Lee}, G.-H., {Lee}, S.-K., \& {Lim}, G. 2019, Nature
  Astronomy, 3, 844

\bibitem[{{Zhang} {et~al.}(2021){Zhang}, {Chen}, \& {Zhao}}]{Zhang_2021}
{Zhang}, H., {Chen}, Y., \& {Zhao}, G. 2021, \apj, 919, 52

\end{thebibliography}

\begin{appendix}
\section{Complementary material}
\label{Appendix}

\begin{table*}
\caption{Properties of the Open Clusters used in this study.}
\label{tab:OCproperties}
\scalebox{0.77}{
\begin{tabular}{lccccccccccccc}

\hline\hline
Cluster & $[$M/H$]$ (dex) & $[$$\alpha$/Fe$]$ (dex) & Age (Gyr)& R$_{\rm GC}$ (pc) & $e$ (kpc) & $Z_{max}$ & V$_R$ & V$_{\phi}$ & |V$_Z$| & $J_{R}$ & $J_{Z}$ & $L_{Z}$ & N \\
\hline
  Alessi 1 & -0.052 & -0.029 & 1.445 & 8637.1 & 0.063 & 0.23 & 17.29 & 230.996 & 12.241 & 5.325 & 1.731 & 1995.561 & 71\\
  Berkeley 89 & 0.047 & -0.024 & 2.089 & 8473.7 & 0.173 & 0.338 & 28.563 & 205.102 & 8.261 & 36.858 & 3.569 & 1753.963 & 177\\
  FSR 0241 & -0.002 & -0.056 & 1.318 & 8189.7 & 0.154 & 0.415 & 15.937 & 266.624 & 3.232 & 50.9 & 3.589 & 2184.002 & 43\\
  FSR 0278 & 0.114 & -0.076 & 2.188 & 8352.8 & 0.165 & 0.313 & -45.211 & 254.632 & 9.444 & 40.024 & 2.276 & 2127.254 & 49\\
  FSR 1378 & -0.045 & 0.006 & 2.239 & 8747.9 & 0.189 & 0.354 & 46.001 & 260.03 & 15.312 & 57.306 & 2.428 & 2276.28 & 130\\
  Hogg 4 & -0.077 & -0.081 & 1.479 & 8466.1 & 0.042 & 0.233 & 3.243 & 230.825 & 16.138 & 2.883 & 1.873 & 1970.592 & 216\\
  IC 4756 & -0.151 & 0.049 & 1.288 & 7878.0 & 0.057 & 0.072 & 11.08 & 229.497 & 2.456 & 4.332 & 0.208 & 1807.62 & 530\\
  LP 1540 & 0.427 & -0.064 & 1.622 & 7683.3 & 0.087 & 0.13 & -19.853 & 253.056 & 5.881 & 9.985 & 0.563 & 1944.274 & 69\\
  LP 5 & -0.066 & 0.003 & 2.692 & 7755.2 & 0.097 & 0.473 & -3.213 & 259.674 & 8.499 & 13.82 & 5.901 & 2013.559 & 335\\
  NGC 2354 & -0.116 & 0.0 & 1.413 & 8956.3 & 0.038 & 0.166 & -6.573 & 232.945 & 6.803 & 2.315 & 0.908 & 2087.446 & 336\\
  NGC 2423 & 0.007 & 0.001 & 1.096 & 8860.4 & 0.097 & 0.088 & -26.474 & 226.895 & 2.033 & 12.815 & 0.251 & 2010.868 & 423\\
  NGC 2477 & 0.073 & -0.08 & 1.122 & 8742.0 & 0.121 & 0.161 & -34.391 & 245.759 & 4.339 & 22.73 & 0.698 & 2149.44 & 2650\\
  NGC 2627 & -0.116 & 0.024 & 1.862 & 8940.0 & 0.054 & 0.285 & -16.129 & 234.607 & 9.453 & 4.086 & 2.411 & 2099.097 & 298\\
  NGC 2682 & 0.045 & -0.041 & 4.266 & 8842.4 & 0.086 & 0.572 & 25.471 & 227.857 & 13.116 & 10.1 & 8.454 & 2015.661 & 1208\\
  NGC 3680 & -0.079 & 0.069 & 2.188 & 8021.8 & 0.021 & 0.326 & -6.284 & 237.62 & 1.904 & 0.567 & 3.733 & 1905.841 & 129\\
  NGC 6791 & 0.136 & 0.025 & 6.31 & 7855.3 & 0.296 & 1.081 & 66.26 & 185.718 & 14.839 & 100.133 & 25.889 & 1482.269 & 1618\\
  NGC 6811 & -0.061 & 0.034 & 1.072 & 8116.7 & 0.121 & 0.309 & -27.764 & 255.925 & 3.593 & 21.635 & 2.48 & 2077.054 & 286\\
  NGC 6819 & 0.034 & -0.038 & 2.239 & 7930.6 & 0.046 & 0.494 & 8.319 & 245.501 & 15.815 & 2.926 & 7.296 & 1947.927 & 1498\\
  NGC 6939 & -0.008 & -0.046 & 1.698 & 8609.5 & 0.062 & 0.486 & -9.561 & 247.334 & 10.224 & 5.697 & 6.062 & 2130.147 & 608\\
  NGC 6940 & -0.016 & -0.025 & 1.349 & 7959.7 & 0.082 & 0.221 & -23.522 & 246.01 & 12.921 & 9.092 & 1.569 & 1958.012 & 556\\
  NGC 6991 & -0.098 & 0.042 & 1.549 & 8242.0 & 0.101 & 0.144 & 32.043 & 237.999 & 9.919 & 13.642 & 0.674 & 1961.587 & 233\\
  NGC 7044 & 0.027 & -0.005 & 1.66 & 8646.8 & 0.15 & 0.29 & -42.87 & 231.424 & 6.567 & 32.226 & 2.198 & 2031.237 & 653\\
  NGC 7762 & -0.107 & -0.009 & 2.042 & 8714.0 & 0.143 & 0.304 & -3.153 & 206.131 & 18.209 & 25.903 & 2.997 & 1796.561 & 413\\
  Pismis 12 & 0.02 & -0.053 & 1.23 & 8564.7 & 0.064 & 0.161 & 16.159 & 230.691 & 4.349 & 5.685 & 0.886 & 1978.272 & 157\\
  Pismis 3 & -0.211 & -0.047 & 3.162 & 8935.9 & 0.095 & 0.154 & 19.555 & 251.02 & 8.82 & 14.649 & 0.619 & 2253.026 & 979\\
  Ruprecht 147 & 0.063 & 0.046 & 3.02 & 7972.9 & 0.174 & 0.257 & -55.673 & 231.803 & 16.762 & 38.937 & 1.947 & 1847.874 & 170\\
  Ruprecht 172 & -0.088 & 0.003 & 1.047 & 7956.9 & 0.112 & 0.096 & 27.337 & 254.583 & 1.81 & 17.412 & 0.278 & 2028.749 & 71\\
  Skiff J1942+38.6 & 0.028 & -0.078 & 1.479 & 7876.7 & 0.078 & 0.349 & -14.319 & 224.608 & 3.3 & 7.463 & 4.309 & 1769.132 & 53\\
  Trumpler 19 & 0.012 & -0.035 & 4.898 & 7757.5 & 0.242 & 0.299 & -70.368 & 261.163 & 14.208 & 84.178 & 1.942 & 2025.851 & 611\\
  UBC 1083 & -0.054 & -0.013 & 1.047 & 7654.9 & 0.142 & 0.047 & 23.544 & 266.514 & 3.245 & 28.197 & 0.073 & 2040.629 & 60\\
  UBC 1116 & -0.012 & -0.001 & 3.548 & 8478.2 & 0.166 & 0.88 & 34.347 & 261.988 & 5.521 & 43.392 & 13.189 & 2230.123 & 53\\
  UBC 141 & -0.155 & 0.02 & 2.089 & 8084.8 & 0.092 & 0.255 & 27.855 & 231.648 & 15.351 & 10.824 & 2.187 & 1872.871 & 38\\
  UBC 1493 & -0.166 & -0.007 & 1.175 & 8774.5 & 0.092 & 0.413 & -25.99 & 228.456 & 0.407 & 11.416 & 4.762 & 2011.218 & 32\\
  UBC 255 & -0.099 & 0.008 & 2.399 & 8191.7 & 0.12 & 0.22 & -15.915 & 261.708 & 0.18 & 21.357 & 1.249 & 2143.968 & 45\\
  UBC 268 & 0.279 & -0.028 & 1.698 & 7728.3 & 0.054 & 0.218 & -9.604 & 229.555 & 5.673 & 3.632 & 1.889 & 1778.43 & 81\\
  UBC 284 & 0.078 & -0.103 & 1.585 & 7548.4 & 0.102 & 0.156 & 4.778 & 262.333 & 11.457 & 14.79 & 0.758 & 1979.781 & 157\\
  UBC 374 & 0.094 & -0.109 & 1.288 & 8049.5 & 0.085 & 0.055 & 14.246 & 241.543 & 1.284 & 12.109 & 0.104 & 1944.26 & 43\\
  UBC 577 & -0.151 & 0.06 & 2.754 & 7709.0 & 0.096 & 0.312 & 27.594 & 247.991 & 15.792 & 11.912 & 3.022 & 1911.476 & 9\\
  UBC 586 & -0.059 & 0.023 & 1.096 & 8285.2 & 0.056 & 0.501 & -9.339 & 246.761 & 13.543 & 4.668 & 6.831 & 2048.356 & 36\\
  UPK 21 & -0.014 & -0.292 & 1.148 & 7704.9 & 0.094 & 0.125 & 23.305 & 252.365 & 8.39 & 11.657 & 0.518 & 1943.996 & 47\\
  UPK 84 & -0.179 & 0.081 & 1.0 & 7904.3 & 0.145 & 0.227 & -39.095 & 255.679 & 0.267 & 29.185 & 1.395 & 2020.677 & 66\\
\hline
\end{tabular}}
\tablefoot{The stellar parameters and abundances of member stars (callibrated metallicity and $[$$\alpha$/Fe$]$) are from the Gaia General Stellar Parametrizer from spectroscopy (GSPspec) \citep{Recio-Blanco_2022} and the membership of stars and  ages are taken from \citet{CantatGaudin20}, while  orbital parameters are recomputed in the present work.}
\end{table*}

\begin{figure}
  \resizebox{\hsize}{!}{\includegraphics{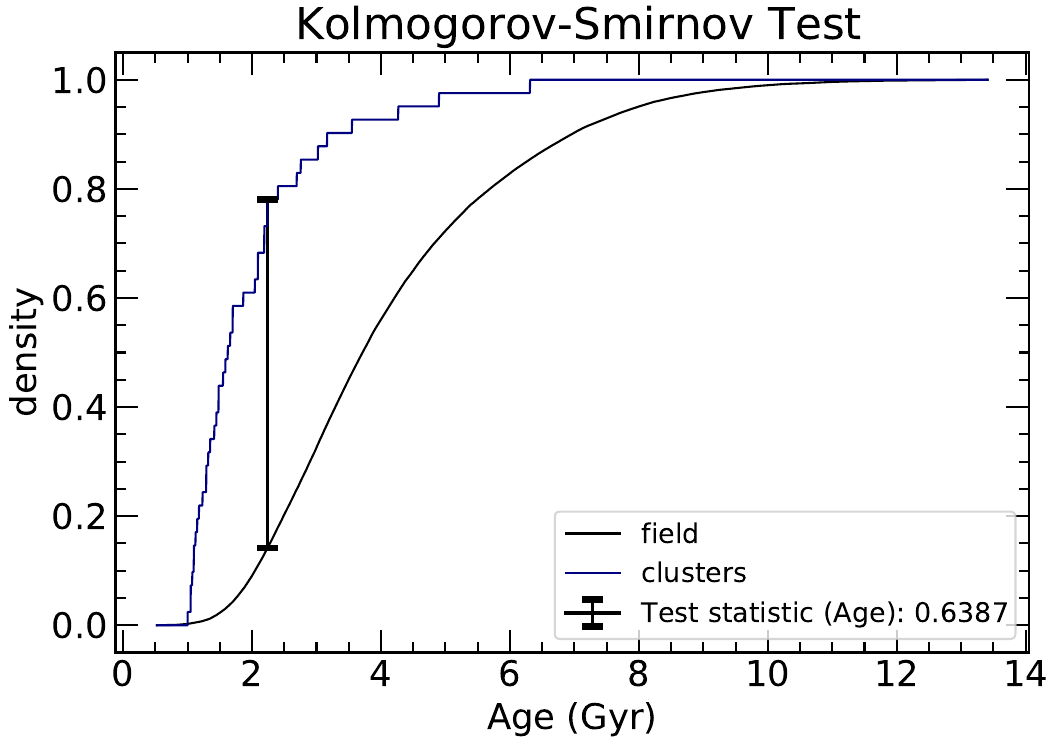}}
  \caption{Kolmogorov-Smirnov (K-S) test comparing the cumulative distributions of Age for the two samples: field (black) and clusters (blues). The vertical bars represents the values of the test statistic, which is the maximum absolute difference between the two cumulative distributions}
  \label{fig:KS_Age}
\end{figure}

\begin{figure}
  \resizebox{\hsize}{!}{\includegraphics{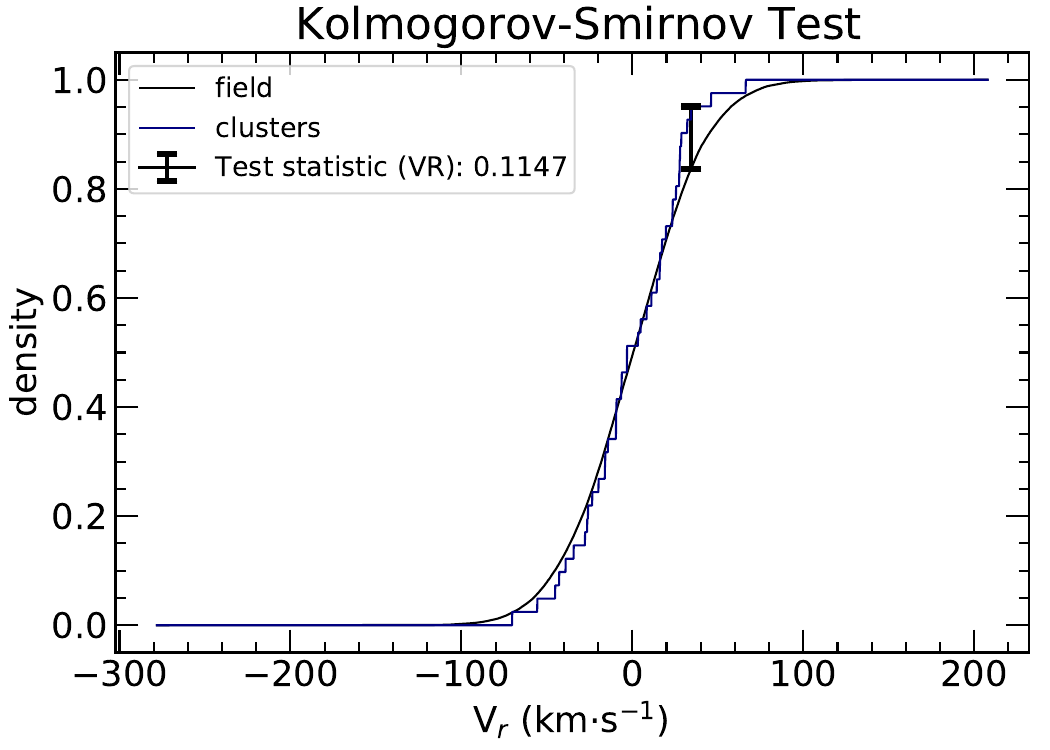}}
  \caption{K-S for V$_r$. Symbols and colours as in Fig.~\ref{fig:KS_Age}.}
  \label{fig:KS_VR}
\end{figure}

\begin{figure}
  \resizebox{\hsize}{!}{\includegraphics{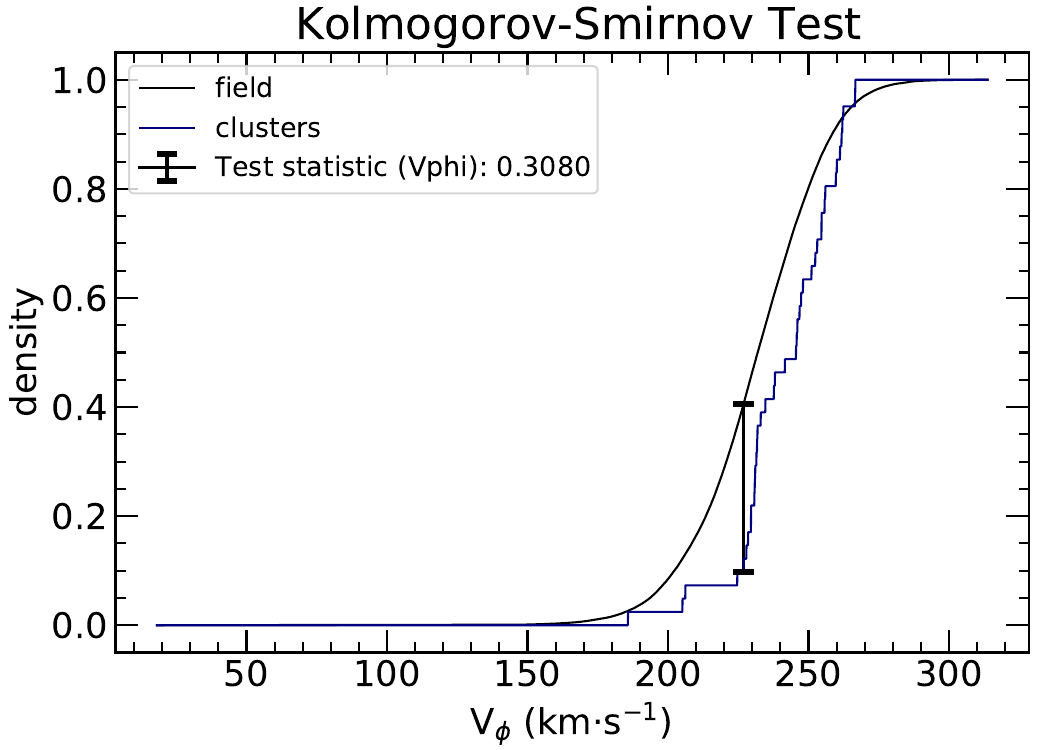}}
  \caption{K-S for V$_{\phi}$. Symbols and colours as in Fig.~\ref{fig:KS_Age}.}
  \label{fig:KS_Vphi}
\end{figure}

\begin{figure}
  \resizebox{\hsize}{!}{\includegraphics{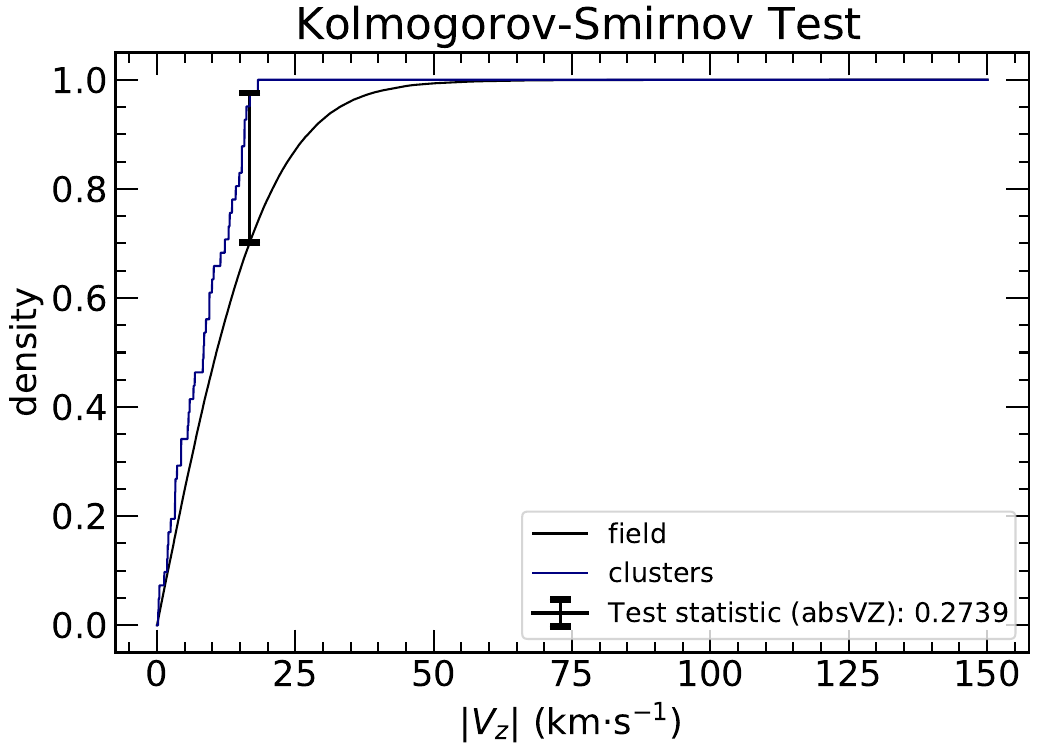}}
 \caption{K-S for |V$_{Z}$|. Symbols and colours as in Fig.~\ref{fig:KS_Age}.}  \label{fig:KS_absVZ}
\end{figure}

\begin{figure}
  \resizebox{\hsize}{!}{\includegraphics{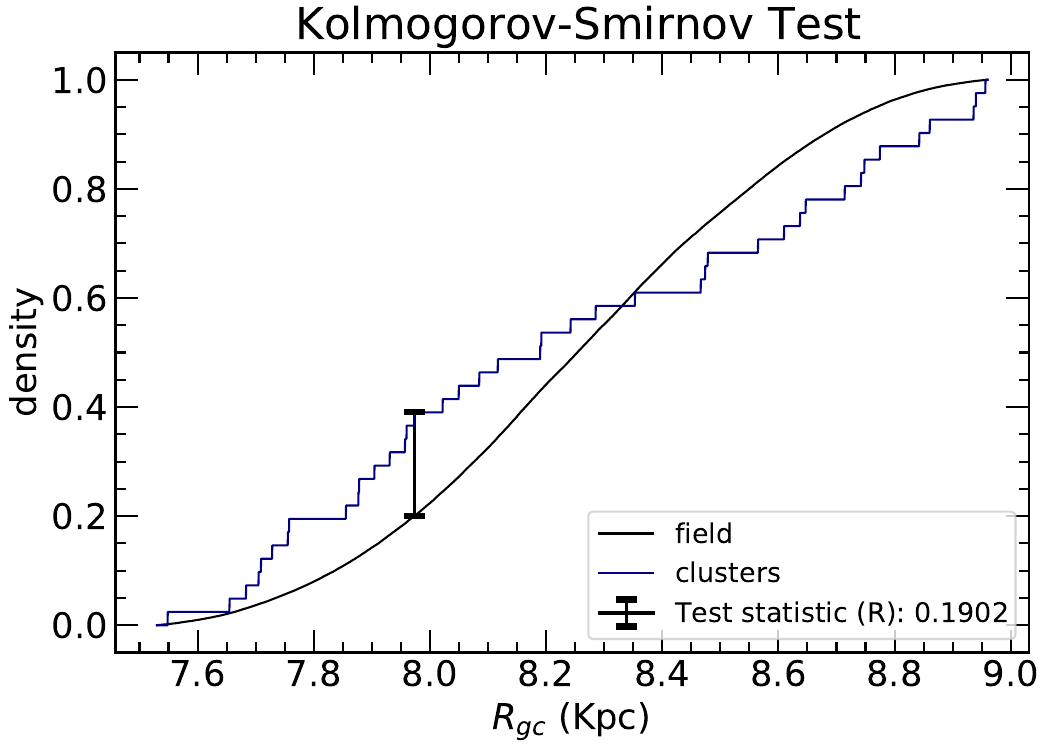}}
 \caption{K-S for R. Symbols and colours as in Fig.~\ref{fig:KS_Age}.}
  \label{fig:KS_R}
\end{figure}

\begin{figure}
  \resizebox{\hsize}{!}{\includegraphics{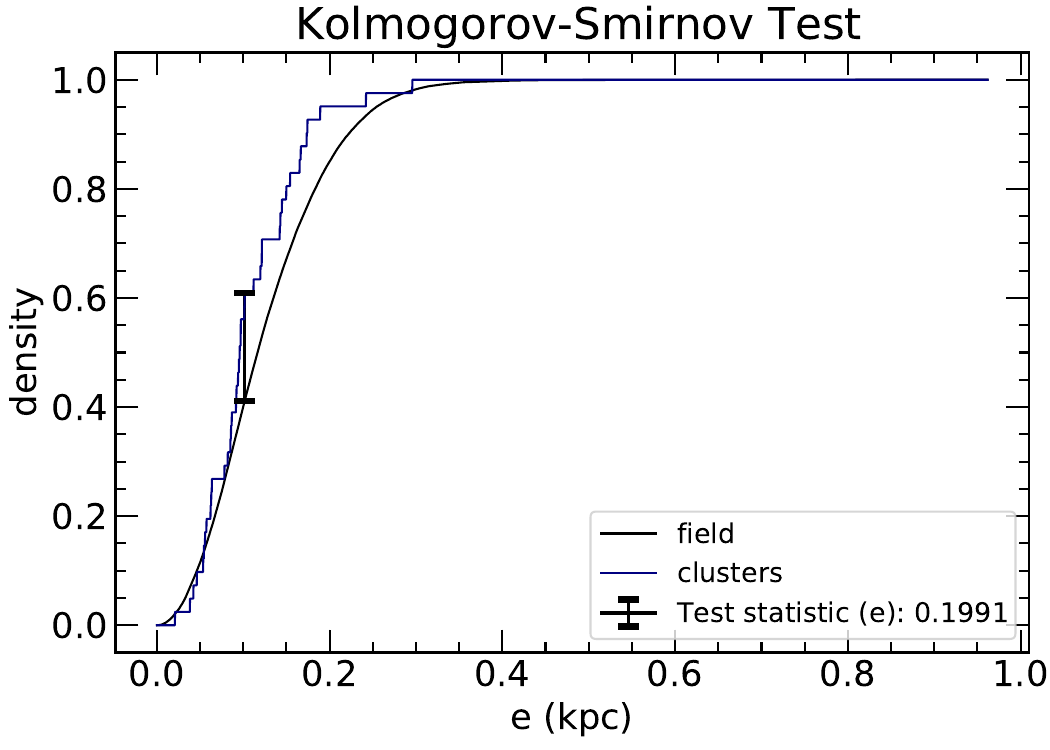}}
 \caption{K-S for $e$. Symbols and colours as in Fig.~\ref{fig:KS_Age}.}
  \label{fig:KS_e}
\end{figure}

\begin{figure}
  \resizebox{\hsize}{!}{\includegraphics{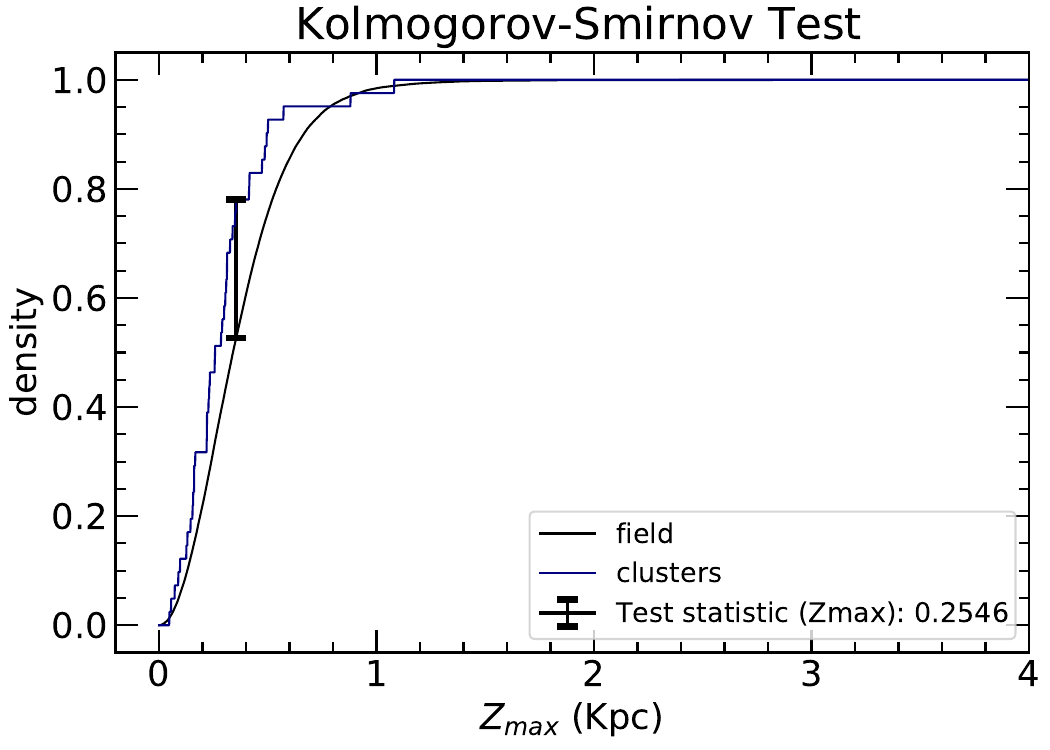}}
 \caption{K-S for Z$_{max}$. Symbols and colours as in Fig.~\ref{fig:KS_Age}.}
  \label{fig:KS_Zmax}
\end{figure}

\begin{figure}
  \resizebox{\hsize}{!}{\includegraphics{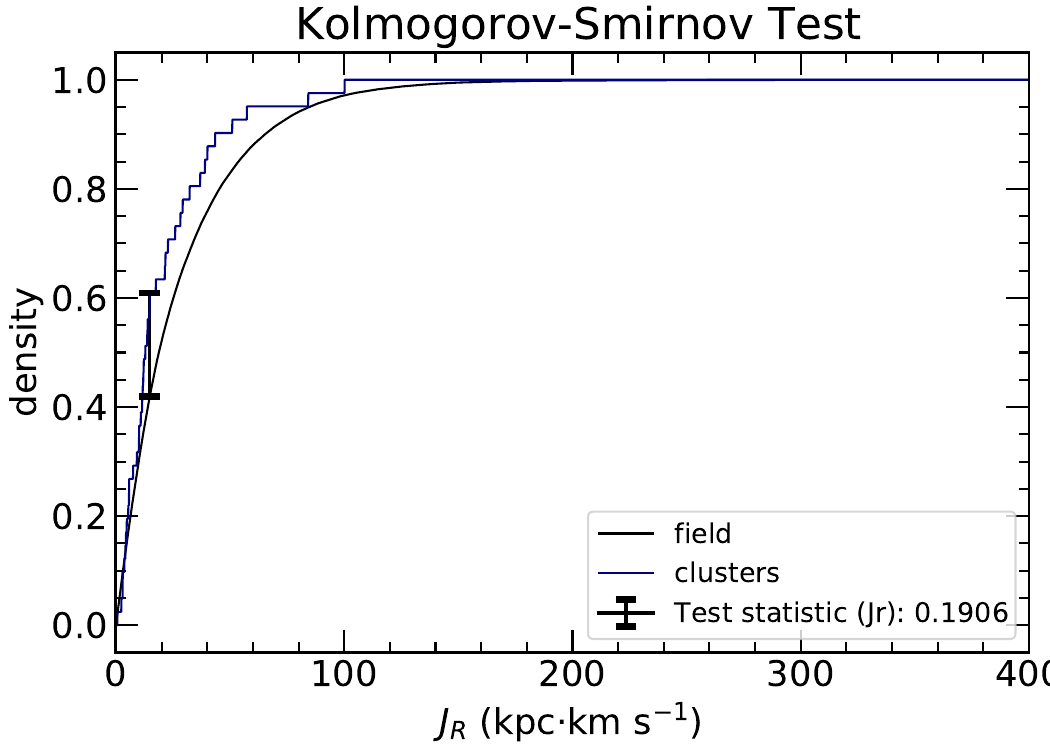}}
 \caption{K-S for J$_{R}$. Symbols and colours as in Fig.~\ref{fig:KS_Age}.}
  \label{fig:KS_Jr}
\end{figure}

\begin{figure}
  \resizebox{\hsize}{!}{\includegraphics{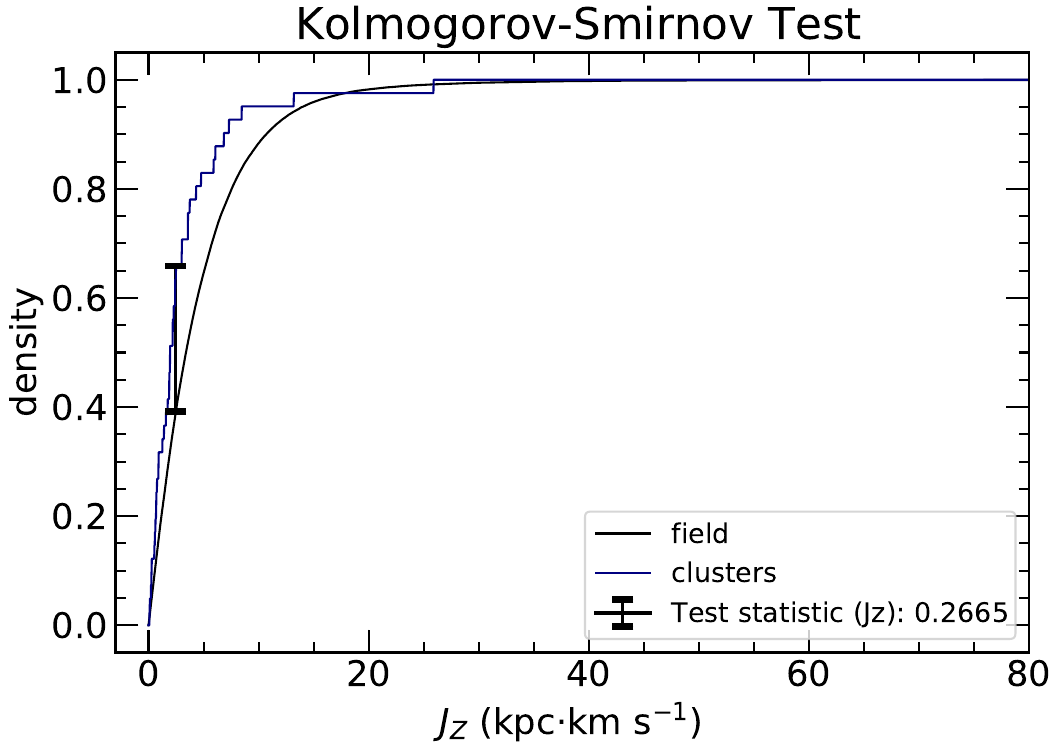}}
 \caption{K-S for J$_{Z}$. Symbols and colours as in Fig.~\ref{fig:KS_Age}.}
  \label{fig:KS_Jz}
\end{figure}

\begin{figure}
  \resizebox{\hsize}{!}{\includegraphics{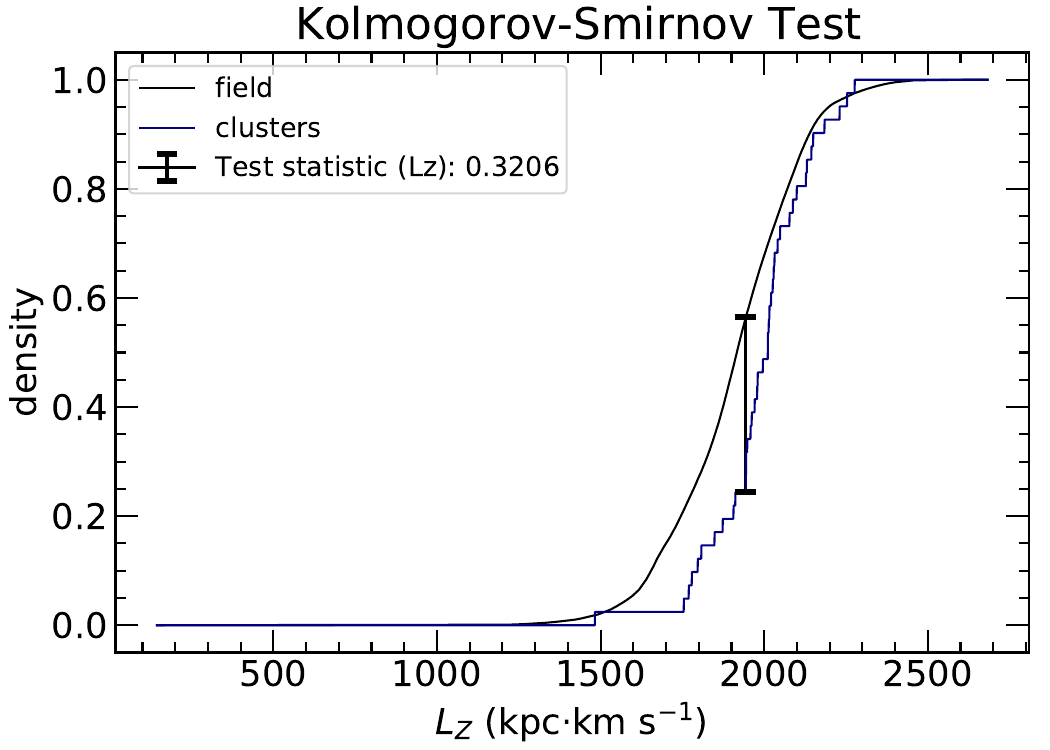}}
 \caption{K-S for L$_{Z}$. Symbols and colours as in Fig.~\ref{fig:KS_Age}.}
  \label{fig:KS_Lz}
\end{figure}

\end{appendix}

\end{document}